\documentclass[twocolumn,useAMS,usenatbib]{mn2e}
\usepackage{graphicx,amsmath}
\usepackage{bm}

\newcommand{\rmd}{{\rm d}}

\newcommand{\rpi}{\Pi}

\newcommand{\beq}{\begin{equation}}
\newcommand{\eeq}{\end{equation}}
\newcommand{\beqa}{\begin{eqnarray}}
\newcommand{\eeqa}{\end{eqnarray}}

\title[Intrinsic alignments in SDSS]
{
Detection of large scale intrinsic 
ellipticity-density correlation from the 
Sloan Digital Sky Survey and implications for weak lensing surveys}

\author[Mandelbaum et al.]
 {Rachel Mandelbaum$^1$\thanks{Electronic address:
    {\tt rmandelb@princeton.edu}},
  Christopher M. Hirata$^{1,2}$,
  Mustapha Ishak$^{3,4}$, 
\newauthor
  Uro\v s Seljak$^{1,5}$, and
  Jonathan Brinkmann$^6$
\\$^1$Department of Physics, Jadwin Hall, Princeton University,
      Princeton, NJ 08544, USA
\\$^2$Institute for Advanced Study, Einstein Drive,
      Princeton, NJ 08540, USA
\\$^3$Department of Astrophysical Sciences, Princeton University,
      Princeton, NJ 08544, USA
\\$^4$Department of Physics, The University of Texas at Dallas,
      Richardson TX 75083, USA
\\$^5$International Centre for Theoretical Physics, Strada Costiera 11,
      34014 Trieste, Italy
\\$^6$Apache Point Observatory, 2001 Apache Point Road,
      Sunspot NM 88349, USA
}

\date{\today}

\begin{document}
\maketitle

\begin{abstract}
The power spectrum of weak lensing shear caused by large-scale structure 
is an emerging tool for precision cosmology, in particular for measuring 
the effects of dark energy on the growth of structure at low redshift.  
One potential source of systematic error is 
intrinsic alignments of ellipticities of neighbouring galaxies (II 
correlation) that could mimic the correlations due to lensing. A related 
possibility pointed out by Hirata and Seljak (2004) 
is correlation between the intrinsic ellipticities of galaxies 
and the density field responsible for gravitational lensing shear (GI 
correlation). We present constraints on both the II and GI
correlations using 265~908 spectroscopic 
galaxies from the Sloan Digital Sky Survey (SDSS), and using galaxies as 
tracers of the mass in the case of the GI analysis.  The availability of 
redshifts in the SDSS allows us to select galaxies at small radial
separations, which both reduces noise in the intrinsic alignment measurement 
and suppresses galaxy-galaxy lensing (which otherwise swamps the GI 
correlation).  While we find no detection of the II correlation, our
results are nonetheless statistically consistent with recent
detections found using the SuperCOSMOS 
survey. Extrapolation of these limits to   
cosmic shear surveys at $z \sim 1$ suggests that the II correlation is
unlikely to have been a significant source of error for current
measurements of $\sigma_8$ with $\sim 10$ per cent accuracy, but may
still be an issue for future surveys with projected statistical errors
below the one per cent level unless eliminated using 
photometric redshifts. In contrast, 
we have a clear detection of GI correlation in galaxies brighter than
$L_*$ that persists to the largest scales probed (60 $h^{-1}$Mpc) and with 
a sign predicted by theoretical models. 
%This could contaminate cosmic shear surveys at a 
%depth of $z\sim 1$ as much as the 50 per cent level depending on the source 
%sample used. 
This correlation could cause the
existing lensing surveys at $ z\sim 1$ to 
underestimate the linear amplitude 
of fluctuations by as much as 20 per cent depending on the source sample used, 
while for surveys at $z \sim 0.5$ the underestimation may reach 30 per cent. 
The GI contamination is dominated by the brightest galaxies, 
%with our brightest subsample showing contamination a factor of $\ge 10$ 
%worse than the faintest sample.  This contamination 
possibly due to the alignment of brightest cluster galaxies (BCGs) with
the cluster ellipticity due to anisotropic infall along filaments, although
other sources of contamination cannot be excluded at this point.  
We propose that cosmic shear surveys should consider rejection of BCGs
from their source catalogs as a  test for GI contamination.
Future high precision weak lensing surveys must develop methods to
search for and remove this contamination if they are to achieve their
promise.  

\end{abstract}


\section{Introduction}

Weak gravitational lensing of distant galaxies has emerged as a powerful 
method for directly measuring the matter distribution in the universe 
(e.g. \citealt{1999ARA&A..37..127M, 2001PhR...340..291B, 
2003ARA&A..41..645R}) following detection of the two-point function of the 
lensing-induced shear by several groups \citep{2000MNRAS.318..625B,
  2000A&A...358...30V, 2001ApJ...552L..85R, 2002ApJ...572...55H, 
2002A&A...393..369V, 2003MNRAS.341..100B, 2003AJ....125.1014J}.  Also, 
weak lensing has been shown to be a very promising tool for precision cosmology 
\citep{2002PhRvD..65b3003H, 2002PhRvD..65f3001H, 2003PhRvL..91d1301A,
  2004PhRvD..70l3515B, 2004ApJ...600...17B, 2004PhRvD..69h3514I,  
2005astro.ph..1594I, 2005PhRvD..71b4026S,
2005A&A...429..383T, 2005PhRvD..72f3501U}, and has the clear advantage
of being the 
simplest of the low-redshift cosmological probes to understand
theoretically, since it is sensitive primarily to the dark matter
whose interactions are described purely by gravity.  Thus, most of the
potential systematic errors in weak lensing are observational,
i.e. associated with attempts to measure a small signal, in contrast
with galaxy surveys, where large amounts of information are ``lost''
on quasilinear or nonlinear scales because of the lack of
a robust prediction. 

Examples of the observational systematics in weak lensing include 
uncertainties in the point-spread function (PSF) of the telescope, errors 
in correcting the ellipticity of a source galaxy for the PSF effects, 
star-galaxy separation, selection biases, deblending errors, detector 
nonlinearities, and noise-related biases.  The list is long, and much of 
the effort by the weak lensing observers in the past several years has 
been devoted to these issues \citep{2000ApJ...537..555K, 
2001A&A...366..717E, 2002AJ....123..583B, 2003MNRAS.343..459H, 
2004ApJ...613L...1V, 2005A&A...429...75V}.  Nevertheless, it is important 
to remember that there are astrophysical uncertainties associated with 
weak lensing. Some of these, such as the 
error in $N$-body simulations or 
baryonic cooling effects, are limited to small scales only and are likely 
to be reduced significantly as the simulations improve  
\citep{2004APh....22..211W}.  
Another astrophysical uncertainty is the redshift distribution $\rmd 
N/\rmd z$ of the source galaxies.  At least at the brighter magnitudes, 
these can be determined via spectroscopic redshifts 
\citep{2004ApJ...600...17B, 2005PhRvD..71b3002I, 2005astro.ph..6614M}, and 
at all magnitudes the distributions can be tested by comparing the 
galaxy-galaxy lensing signal among different source samples 
\citep{2005MNRAS.361.1287M, 2005astro.ph..6030H}.  Future low-frequency 
radio surveys may also open up the possibility of obtaining redshifts on 
many galaxies from the H$\,${\sc i} 21$\;$cm line (e.g. 
\citealt{2004NewAR..48.1013R}).  In short, while the 
redshift distribution problem is not solved, there is no fundamental 
impediment to an accurate determination of $\rmd N/\rmd 
z$.

The final major astrophysical contaminant of weak lensing may be intrinsic 
alignments, i.e. correlations of the galaxy ellipticites with each other 
or with the density field.  These correlations violate the usual 
assumption in lensing shear surveys that the source galaxies are randomly 
oriented so that any correlation between the apparent ellipticities of 
distinct objects is due to lensing.  The purpose of this paper is to 
provide an observational constraint on these types of intrinsic 
alignments, and then to assess the implied contamination of upcoming 
cosmic shear surveys.

There are two types of intrinsic alignments that can contaminate the 
cosmic shear power spectrum, namely the intrinsic ellipticity-intrinsic 
ellipticity (II) correlation and the gravitational shear-intrinsic 
ellipticity (GI) correlation.  The II contamination is the easier to 
understand: it results from the possibility that two source galaxies are 
physically near each other and have correlated intrinsic ellipticities.  
This effect contributes an additive bias to any two-point shear
statistic, though as pointed out by \cite{2002A&A...396..411K} and
\cite{2003MNRAS.339..711H}, its effects   
can be minimized by using photometric redshifts to cross-correlate the
shapes of galaxies at different redshifts.  The 
GI contamination results from the possibility that, given two source 
galaxies, one is in front of the other.  In this case, the intrinsic 
ellipticity of the nearby galaxy may be correlated with the density field 
that lenses the more distant galaxy, thus yielding a spurious contribution 
to the shear two-point function \citep{2004PhRvD..70f3526H}.
This contribution cannot be eliminated by selecting galaxies at
different redshifts as for the II contamination, and in fact doing so
will tend to increase any GI contamination.

So far, the main methods used to estimate the intrinsic alignments and 
assess their implications for cosmic shear have been theoretical results 
(analytical or simulation-based) and observations of low-redshift galaxies 
for which cosmic shear is negligible and any observed correlations must be 
intrinsic.  Both methods have their limitations: the theory of galaxy 
alignments is subject to all of the uncertainties involved in the theory 
of galaxy formation, whereas the observations must be extrapolated from 
observable redshifts $z\sim 0.1$ to the redshifts $z\sim 1$ of the source 
galaxies used for lensing.  Nevertheless, the theoretical predictions 
vary by more than an order of magnitude for II 
\citep{2000ApJ...545..561C, 2000MNRAS.319..649H, 2000ApJ...532L...5L, 
2001ApJ...555..106L, 2001MNRAS.320L...7C, 2001ApJ...559..552C, 
2002MNRAS.335L..89J}, so there is a clear role for observations in 
distinguishing which of these models is correct.  The theory of GI 
correlations is even more rudimentary: only simple analytical models exist 
\citep{2002astro.ph..5212H, 2004PhRvD..70f3526H}.

At present, the most statistically powerful constraint on II correlations 
comes from the SuperCOSMOS data \citep{2002MNRAS.333..501B}, which were 
re-analyzed by \citet{2004MNRAS.347..895H} with the conclusion that the II 
correlations are smaller than most of the theoretical predictions.  The Sloan 
Digital Sky Survey (SDSS) is ideally suited to improving upon SuperCOSMOS 
because of the availability of spectroscopic redshifts, which allows for
measurement of the correlations only between galaxies that are near
each other in  
three-dimensional space, without the ``noise'' introduced by pairs of 
physically unassociated galaxies that happen to lie along the same line of 
sight.  One can also attempt to constrain the GI correlation using 
low-redshift surveys such as SDSS.  The GI correlation is essentially a 
measure of the correlation between the intrinsic ellipticities and the 
matter density; on sufficiently large scales, one can use the correlation 
between the intrinsic ellipticities and the galaxy density as a proxy.  
In this case the measurement of the GI correlation is model-dependent;  
however, we will argue that on large scales the associated uncertainty is 
probably less than the uncertainty associated with extrapolation to high 
redshift.  The ability of spectroscopic redshifts to isolate pairs of 
galaxies at the same redshift is valuable for GI correlation studies to 
reduce noise, just as for II.  It also enables us to cleanly separate GI 
from the ``spurious'' (for this work!) density-ellipticity correlations 
caused by galaxy-galaxy lensing.

We note that there have been several observational studies of intrinsic 
alignments that were not motivated by potential contamination of cosmic 
shear measurements.  These include \citet{2002AJ....124..733B} and 
\citet{2004MNRAS.353..529H}, who were interested in intrinsic alignment 
contamination of galaxy-galaxy lensing; \citet{2000ApJ...543L.107P}, 
\citet{2001ApJ...555..106L}, and \citet{2002ApJ...567L.111L}, who were 
interested in using the tidal influence on intrinsic alignments to 
reconstruct the matter density field; and \citet{2004ApJ...613L..41N}, who 
were interested in the formation of disk galaxies.  These results are 
difficult (or, in some cases, impossible) to interpret in the context of 
cosmic shear because they are made at too small a physical scale, use 
measures of the galaxy shape not easily related to ellipticity, measure 
higher-order moments instead of density-ellipticity or 
ellipticity-ellipticity two-point functions, have complicated selection 
criteria that differ dramatically from those relevant to cosmic shear, or 
do not provide sufficient characterization of statistical errors.

The outline of the paper is as follows.  We begin by introducing the formalism used in this paper in \S\ref{S:formalism}.   The SDSS 
data used for this analysis is described in \S\ref{S:data}.  We
describe the calculation of the relevant correlation functions and
show the resulting signal in \S\ref{S:corrfun}.  In  
\S\ref{S:contam} we use the computed signal to derive estimates of the 
contamination due to intrinsic alignments in current and future surveys.  
After proposing an explanation for the detected signal in \S\ref{S:BCG}, 
we conclude with a discussion of the implications of our results, and 
suggestions for future work.

Here we note the cosmological model and units adopted for this work.
Pair separations are measured in comoving $h^{-1}$Mpc, with the
angular diameter distance computed in a flat $\Lambda$CDM cosmology
with $\Omega_m=0.3$, $\Omega_{\Lambda}=0.7$.

\section{Formalism}\label{S:formalism}

The formalism for the analysis of the lensing shear two-point function 
\citep{1991ApJ...380....1M} and the intrinsic alignment contamination is 
well developed.  We will briefly summarize the important equations here, 
and then define some new variables that relate to observables in galaxy 
surveys.  Our notation is consistent with that of 
\citet{2004PhRvD..70f3526H}.

The observed shear $\bgamma$ of a galaxy is a sum of two components: the 
gravitational lensing-induced shear $\bgamma^G$, and the ``intrinsic 
shear'' $\bgamma^I$, which includes any non-lensing shear, typically
due to local tidal fields.  Therefore the $E$-mode shear power
spectrum between  
any two redshift bins $\alpha$ and $\beta$ is the sum of the gravitational 
lensing power spectrum (GG), the intrinsic-intrinsic, and the 
gravitational-intrinsic terms,
\beq
C_L^{EE}(\alpha\beta) =
C_L^{EE,GG}(\alpha\beta)+
C_L^{EE,II}(\alpha\beta)+
C_L^{EE,GI}(\alpha\beta).
\eeq
The pure gravitational lensing term is given by the Limber integral
\beq
C_L^{EE,GG}(\alpha\beta) = \int \frac{W_\alpha(\chi) 
W_\beta(\chi)}{\sin_K^2\chi}P_\delta(L\csc_K\chi,\,\chi)\,\rmd\chi,
\label{eq:cegg}
\eeq
where $\chi$ is the comoving distance of the lensing screen, $P_\delta$ is 
the matter power spectrum, and the modified trigonometric functions
are defined by
\begin{equation}
\sin_K\chi = 
\begin{cases}
R_0 \sinh\left(\frac{\chi}{R_0}\right), &\Omega_K>0,
R_0=\frac{c}{H_0\sqrt{\Omega_K}} \\
\chi, &\Omega_K=0\\
R_0 \sin\left(\frac{\chi}{R_0}\right), &\Omega_K<0, R_0=\frac{c}{H_0\sqrt{|\Omega_K|}}
\end{cases}
\end{equation}
and analogously for $\cos_K\chi$.
The window function for redshift bin $\alpha$ is
\beqa
W_\alpha(\chi) &=& \frac{3}{2}\Omega_mH_0^2(1+z)\sin_K^2\chi
\nonumber \\ && \times
\int f_\alpha(\chi') \left( \cot_K\chi-\cot_K\chi'
\right)\,\rmd\chi'.
\eeqa
Here $f_\alpha(\chi')=\rmd n_\alpha/\rmd\chi'$ is the distribution of 
comoving source distances for source sample $\alpha$, and is equivalent to 
the redshift distribution if one fixes the homogeneous cosmology 
$\chi(z)$.  The intrinsic alignment power spectrum is
\beq
C_L^{EE,II}(\alpha\beta) = \int
\frac{f_\alpha(\chi)f_\beta(\chi)}{\sin_K^2\chi} 
P^{EE}_{\tilde\bgamma^I}(L\csc_K\chi,\,\chi)\,\rmd\chi.
\label{eq:ceii}
\eeq
Here we have introduced the density-weighted intrinsic shear 
$\tilde\bgamma^I=(1+\delta_g)\bgamma^I$, where $\delta_g$ is the galaxy 
(not matter!) overdensity $\rho_g/\overline{\rho}_g-1$, and its
$E$-mode power spectrum  
$P^{EE}_{\tilde\bgamma^I}$.  The GI or ``interference'' term is
\beqa
C_L^{EE,GI}(\alpha\beta) &=& \int \frac{W_\alpha(\chi)f_\beta(\chi)}
{\sin_K^2\chi} P_{\delta,\tilde\bgamma^I}(L\csc_K\chi)\,\rmd\chi
\nonumber \\ && + (\alpha\leftrightarrow\beta).
\label{eq:cegi}
\eeqa
To first order in the deflection angle, there is no $B$-mode induced by 
gravitational lensing shear.  In this case, the $B$-mode power spectrum
contains only a contribution from the intrinsic alignments,
\beqa
C_L^{BB}(\alpha\beta) &=&
C_L^{BB,II}(\alpha\beta)
\nonumber \\ &=&
\int
\frac{f_\alpha(\chi)f_\beta(\chi)}{\sin_K^2\chi}
P^{BB}_{\tilde\bgamma^I}(L\csc_K\chi,\,\chi)\,\rmd\chi.
\label{eq:cbii}
\eeqa
There are lensing-induced contributions to the $B$-mode shear from other 
effects, such as multiple deflections \citep{2002ApJ...574...19C, 
2003PhRvD..68h3002H, 2005PhRvD..71l3527C} and, on small scales, modulation 
of the effective source redshift by galaxy clustering 
\citep{2002A&A...389..729S}.

\citet{2004PhRvD..70f3526H} provided formulas relating the power spectra 
in Eqs.~(\ref{eq:cegg})--(\ref{eq:cbii}) to the real-space correlation 
functions.  These formulas will be extremely useful because we directly 
measure the correlation functions in SDSS.\footnote{We chose to measure 
the correlation function rather than the power spectrum simply because 
this involved minimal modification of pre-existing and well-tested code.}  
If one chooses any two points in the SDSS survey, their separation in 
redshift space can then be identified by the transverse separation 
$r_p$ and the radial redshift space separation $\rpi$.\footnote{The 
redshift space separation is frequently denoted $\pi$; we use $\rpi$ to 
avoid confusion since the number $\pi=3.14...$ appears frequently in this 
paper.}  The $+$ and $\times$ components of the shear are measured with 
respect to the axis connecting the two galaxies (i.e. positive $+$ shear 
is radial, whereas negative $+$ shear is tangential).  Then one can write
the density-intrinsic shear correlation as
\beq
P_{\delta,\tilde\bgamma^I}(k) = -2\pi \int \xi_{\delta +}(r_p,\rpi)
J_2(kr_p)
\,r_p\,\rmd r_p\,\rmd\rpi,
\label{eq:j2}
\eeq
where $\xi_{\delta +}(r_p,\rpi)$ is the correlation function between 
$\delta$ and $\tilde\gamma^I_+$.  The intrinsic-intrinsic correlations are
\beqa
P^{EE}_{\tilde\bgamma^I}(k) &=& \int [
\xi_{++}(r_p,\rpi) J_+(kr_p) + \xi_{\times\times}(r_p,\rpi) J_-(kr_p)]
\nonumber \\ && \times \;
2\pi r_p\,\rmd r_p\,\rmd\rpi
\label{eq:je}
\eeqa
and
\beqa
P^{BB}_{\tilde\bgamma^I}(k) &=& \int [
\xi_{++}(r_p,\rpi) J_-(kr_p) + \xi_{\times\times}(r_p,\rpi) J_+(kr_p)]
\nonumber \\ && \times \;
2\pi r_p\,\rmd r_p\,\rmd\rpi,
\label{eq:jb}
\eeqa
where $J_\pm(x)\equiv[J_0(x)\pm J_4(x)]/2$, and $\xi_{++}$ and 
$\xi_{\times\times}$ represent correlation functions of $\tilde\bgamma_I$.  
The existence of the $\rpi$ integral suggests the definitions
\begin{equation}
w_{\delta+}(r_p) = \int_{-\infty}^{+\infty} 
\xi_{\delta+}(r_p,\rpi)\,\rmd\rpi,
\end{equation}
and similarly for $w_{++}$ and $w_{\times\times}$.

\section{Data}\label{S:data}

The data used here are obtained from the SDSS.  The SDSS 
\citep{2000AJ....120.1579Y} is an ongoing survey to image approximately 
$\pi$ steradians of the sky, and follow up approximately one million of 
the detected objects spectroscopically
\citep{2001AJ....122.2267E,2002AJ....124.1810S,  
2002AJ....123.2945R}. The imaging is carried out by drift-scanning the sky 
in photometric conditions \citep{2001AJ....122.2129H, 
2004AN....325..583I}, in five bands ($ugriz$) \citep{1996AJ....111.1748F, 
2002AJ....123.2121S} using a specially designed wide-field camera 
\citep{1998AJ....116.3040G}. These imaging data are the source of the LSS 
sample that we use in this paper. In addition, objects are targeted for 
spectroscopy using these data \citep{2003AJ....125.2276B} and are observed 
with a 640-fiber spectrograph on the same telescope
\citep{telescope}. All of these data are  
processed by completely automated pipelines that detect and measure 
photometric properties of objects, and astrometrically calibrate the data 
\citep{2001adass..10..269L, 2003AJ....125.1559P,mtpipeline}. The SDSS is well 
underway, and has had five major data releases \citep{2002AJ....123..485S, 
2003AJ....126.2081A, 2004AJ....128..502A, 2005AJ....129.1755A, 
2004AJ....128.2577F, 2005astro.ph..7711A}.  This analysis uses the 
spectroscopically observed galaxies in the Value-Added Galaxy Catalog, 
LSS sample 14 (VAGC; \citealt{2005AJ....129.2562B}), comprising 3423 square 
degrees with SDSS spectroscopic coverage.

\subsection{Galaxy subsamples}

In order to determine the correlation functions $\xi_{\delta +}$, 
$\xi_{++}$, and $\xi_{\times\times}$, one needs a sample of galaxies with 
which to measure the intrinsic shear, and a sample of galaxies with which 
to trace the density field.  For this paper, we use only the SDSS 
spectroscopic galaxies, divided further into luminosity 
subsamples by absolute magnitude in the $r$ filter.  The four subsamples 
are described in Table~\ref{tab:lum}.  These are the same as the L3--L6 
subsamples used in the weak lensing analysis of 
\citet{2005PhRvD..71d3511S}, and are very similar to the L4--L7 subsamples 
used in the galaxy power spectrum analysis of 
\citet{2004ApJ...606..702T}.\footnote{The \citet{2004ApJ...606..702T} 
sample differs in that a redshift cut was imposed to make the sample 
volume-limited, which simplified their analysis.  Also, the 
\citet{2004ApJ...606..702T} subsample numbering is different: their L4 is 
similar to our L3, their L5 to our L4, etc.  Finally, both the
\citet{2004ApJ...606..702T} and the \citet{2005PhRvD..71d3511S}
samples included a slightly stricter apparent magnitude cut than the
sample in this work.} The absolute magnitude cuts 
are defined in terms of $h=H_0/(100\,$km$\,$s$^{-1}\,$Mpc$^{-1})$ such 
that one can implement the cuts without specifying the value of $H_0$.  
The magnitudes have been corrected for extinction, $K$-correction, and 
evolution.  The extinction correction uses the \citet{1998ApJ...500..525S} 
dust reddening maps, with the extinction-to-reddening ratios given in 
\citet{2002AJ....123..485S}.  The $K$-correction to $z=0.1$ is performed using the 
{\sc kcorrect v1\_11} software as described by 
\citet{2003AJ....125.2348B}. The luminosities are corrected for passive evolution to 
$z=0.1$ assuming constant $\rmd M_r/\rmd z=-1.6$, as in 
\citet{2003AJ....125.2276B}. We exclude galaxies lying inside the bright 
star mask.  Random catalogs were generated taking into account the 
variation of spectroscopic completeness with position. The random points 
were assigned absolute magnitudes drawn from a distribution consistent 
with the real sample, and random redshifts were assigned for a given 
magnitude given the selection function $\phi(M | z)$ from 
\citet{2003ApJ...592..819B}.  The sample selection criteria in this work and in
that one are sufficiently similar that when $\phi(M|z)$ from that work is used
to generate random redshifts, their probability distribution matches
the redshift distribution of the real lenses used in this work (within
the noise).  Shape measurements were obtained via  
re-Gaussianization for 96 per cent of this sample (see \S\ref{ss:e}); the 
remainder failed due various problems, such as interpolated or saturated 
centers.

\begin{table*}
\caption{\label{tab:lum}The luminosity subsamples used in this analysis.
The number of galaxies in each sample does not include failures.  The 
linear bias is from \citet{2004ApJ...606..702T} and 
\citet{2005PhRvD..71d3511S}, normalized to $\sigma_8=0.9$.  The 
``total'' linear bias and $\langle z\rangle$ are the number-weighted 
averages of the other samples.}
\begin{tabular}{lcrrcr}
\hline\hline
Subsample & Magnitude & Number of & Number of & Linear
 & $\langle z\rangle$ \\
 & range & galaxies & failures & bias & \\
\hline
L3 & $-20\le M_r+5\log_{10}h< -19$ &  66 312 &  2 723 & 0.85 & 0.07 \\
L4 & $-21\le M_r+5\log_{10}h< -20$ & 118 618 &  4 601 & 0.94 & 0.11 \\
L5 & $-22\le M_r+5\log_{10}h< -21$ &  73 041 &  2 829 & 1.08 & 0.15 \\
L6 & $-23\le M_r+5\log_{10}h< -22$ &   7 937 &    307 & 1.59 & 0.21 \\
\hline
Total &                            & 265 908 & 10 460 & 0.98 & 0.12 \\
\hline\hline
\end{tabular}
\end{table*}

\subsection{Ellipticity data}
\label{ss:e}

In addition to a sample of galaxies, we also need their ellipticities.  
For this purpose, we use the measurements by \citet{2005MNRAS.361.1287M}, who 
obtained shapes for more than 30 million galaxies in the SDSS imaging data 
down to magnitude $r=21.8$ (i.e. far fainter than the spectroscopic limit 
of the SDSS).  This section briefly describes the REGLENS pipeline presented 
in \citet{2005MNRAS.361.1287M}, 
and the one modification we made for this paper.

The REGLENS pipeline obtains galaxy images in the $r$ 
and $i$ filters from the SDSS ``atlas images'' 
\citep{2002AJ....123..485S}.  The basic principle of shear measurement 
using these images is to fit a Gaussian profile with elliptical isophotes 
to the image, and define the components of the ellipticity
\beq
(e_+,e_\times) = \frac{1-(b/a)^2}{1+(b/a)^2}(\cos 2\phi, \sin 2\phi),
\label{eq:e}
\eeq
where $b/a$ is the axis ratio and $\phi$ is the position angle of the 
major axis.  The ellipticity is then an estimator for the shear,
\beq
(\gamma_+,\gamma_\times) = \frac{1}{2\cal R}
\langle(e_+,e_\times)\rangle,
\eeq
where ${\cal R}\approx 0.87$ is called the ``shear responsivity'' and 
represents the response of the ellipticity (Eq.~\ref{eq:e}) to a small 
shear \citep{1995ApJ...449..460K, 2002AJ....123..583B}.  In practice, a 
number of corrections need to be applied to obtain the ellipticity.  The 
most important of these is the correction for the smearing and 
circularization of the galactic images by the PSF; 
\citet{2005MNRAS.361.1287M} uses the PSF maps obtained from stellar images 
by the {\sc psp} pipeline \citep{2001adass..10..269L}, and corrects
for these 
using the re-Gaussianization technique of \citet{2003MNRAS.343..459H},
which includes corrections for non-Gaussianity of both the galaxy
profile and the PSF.  A 
smaller correction is for the optical distortions in the telescope: 
ideally the mapping from the sky to the CCD is shape-preserving 
(conformal), but in reality this is not the case, resulting in a nonzero 
``camera shear.'' In the SDSS, this is a small effect (of order 0.1
per cent) which can be identified and removed using the astrometric solution 
\citep{2003AJ....125.1559P}.  Finally, a variety of systematics tests are 
necessary to determine that the shear responsivity ${\cal R}$ has in fact 
been determined correctly.  We refer the interested reader to 
\citet{2005MNRAS.361.1287M} for the details of these corrections and 
tests.

One modification to the \citet{2005MNRAS.361.1287M} pipeline is necessary 
for the analysis here.  \citet{2005MNRAS.361.1287M} rejected all galaxies 
with {\sc photo}'s CR or INTERP flags set, i.e. if {\sc photo} had to 
interpolate over a cosmic ray, bad CCD column, or bleed trail in order to 
obtain the galaxy image.  These cuts are reasonable for the faint galaxies 
used as sources in galaxy-galaxy lensing; however, the spectroscopic 
galaxies typically have large area and high probability of having one of 
these defects present somewhere within the image.  Indeed, 33 per cent of 
the spectroscopic galaxies are rejected by the \citet{2005MNRAS.361.1287M} 
cuts, and of this subset, 88 per cent come from CR and INTERP.  
Moreover, signal-to-noise is not an issue for the spectroscopic galaxies, 
with typical detections of $\sim 100\sigma$ (versus $\sim 
10$--$20\sigma$ for the photometric galaxies at the limit $r=21.8$), so 
one does not worry about noise-related problems when doing the 
interpolation.  We therefore replaced this cut with a cut on 
INTERP\_CENTER, eliminating only those for which the defect was very
close to the object's center and therefore may have had a relatively larger
effect on the determination of the centroid or other parameters.

\section{Computation of correlation functions}\label{S:corrfun}

\subsection{Estimator}

Our estimator for the the intrinsic alignment correlation function is
a generalization of the  
LS \citep{1993ApJ...412...64L} estimator for the galaxy correlation 
function.  The LS estimator for the galaxy correlation function 
$\xi(r_p,\rpi)$ is
\beq
\hat\xi(r_p,\rpi) = \frac{(D-R)^2}{RR} = \frac{DD-2DR+RR}{RR},
\label{eq:lsxi}
\eeq
where $DD$ is the number of pairs of real (``data'') galaxies with 
separations $r_p$ and $\rpi$, $RR$ is the number of such pairs in a random 
catalog, and $DR$ is the number of pairs of data and random galaxies with 
this separation.  ($DR$ and $RR$ are understood to be rescaled in 
proportion to $N_{\rm gal}$ and $N_{\rm gal}^2$ respectively if the number 
of random catalog galaxies differs from the number of data galaxies, which 
is usually the case.)  From our perspective, the key advantage of the 
random catalog subtraction in Eq.~(\ref{eq:lsxi}) is that $\langle 
D-R\rangle=0$.  This means that if there is any additive systematic error 
$\delta D$ in the data, we find that the bias in the correlation function 
is
\beq
\langle\delta\hat\xi(r_p,\rpi)\rangle
 = \frac{2\langle D-R\rangle\delta D + \delta D^2}{RR};
\label{eq:nobias}
\eeq
since $\langle D-R\rangle=0$, the bias is thus second-order in any 
systematic.  This feature makes the LS estimator more robust than previous 
estimators such as $DD/RR-1$, and is retained in most of the recent 
estimators for the correlation function or power spectrum.

When we compute the galaxy-intrinsic shear correlation function 
$\xi_{g+}(r_p,\rpi)$, we can generalize Eq.~(\ref{eq:lsxi}) to
\beq
\hat\xi_{g+}(r_p,\rpi) = \frac{S_+(D-R)}{RR} = \frac{S_+D-S_+R}{RR},
\label{eq:lsxids}
\eeq
where $S_+D$ is the sum over all pairs with separations $r_p$ and $\rpi$ 
of the $+$ component of shear:
\beq
S_+D = \sum_{i\neq j| r_p,\rpi} \frac{e_+(j|i)}{2\cal R}, 
\eeq
where $e_+(j|i)$ is the $+$ component of the ellipticity of galaxy $j$ 
measured relative to the direction to galaxy $i$, and ${\cal R}$ is the 
shear responsivity.  $S_+R$ is
defined by a similar equation.  Averaged  
over a statistical ensemble, $\langle S_+\rangle=\langle D-R\rangle=0$, so 
the cancellation of systematics to first order (Eq.~\ref{eq:nobias}) also 
applies here.  The subtraction of $S_+R$ in Eq.~(\ref{eq:lsxids}) is 
identical with the usual random catalog subtraction procedure in 
galaxy-galaxy lensing studies \citep{2004AJ....127.2544S, 
2005MNRAS.361.1287M}.  We emphasize that positive $\xi_{g+}$ indicates a
tendency to point towards overdensities of galaxies (i.e., radial
alignment, the opposite of the convention in galaxy-galaxy lensing
that positive shear indicates tangential alignment).  

For the intrinsic shear-shear correlation functions $\xi_{++}(r_p,\rpi)$ 
and $\xi_{\times\times}(r_p,\rpi)$, we simply use the estimators
\beq
\hat\xi_{++} = \frac{S_+S_+}{RR} {\rm ~~and~~}
\hat\xi_{\times\times} = \frac{S_\times S_\times}{RR},
\label{eq:lsxiss}
\eeq
where
\beq
S_+S_+ = \sum_{i\neq j| r_p,\rpi} \frac{e_+(j|i)e_+(i|j)}{(2{\cal R})^2},
\eeq
and similarly for $S_\times S_\times$.  Since $\langle S_+\rangle = 
\langle S_\times\rangle= 0$, the cancellation of systematics to first 
order works again, i.e. the square of any spurious source of shear adds to
Eq.~(\ref{eq:lsxiss}) instead of the shear itself.

\begin{figure}
\includegraphics[angle=-90,width=3.2in]{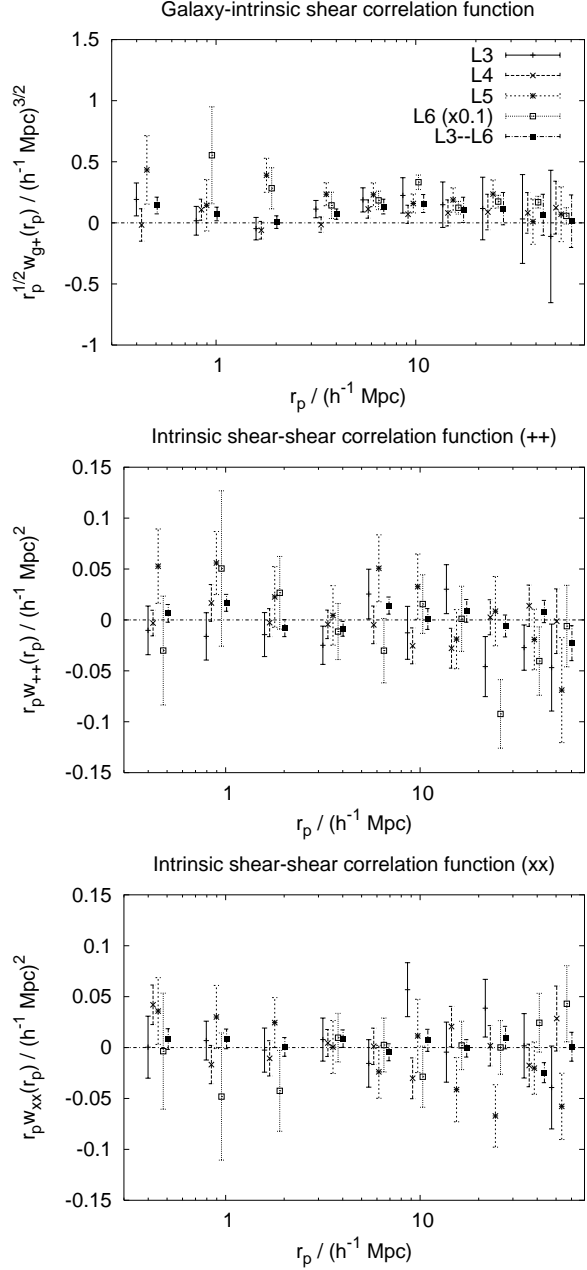}
\caption{\label{fig:rachel}The correlation 
functions $w_{g+}(r_p)$, $w_{++}(r_p)$, and $w_{\times\times}(r_p)$ 
obtained from the L3, L4, L5, L6, and the full galaxy samples.  Each
of the 10 bins contains the same range in $r_p$ for each of the samples, but 
some of the error bars have been slightly displaced horizontally for 
clarity (L5 has not been displaced).  The L6 data have been multiplied by 
0.1 so that they can fit on the same scale. All the errors are 68 per cent
confidence bands, and the errorbars are highly correlated on large scales.} 
\end{figure}

\begin{figure}
\includegraphics[angle=-90,width=3.2in]{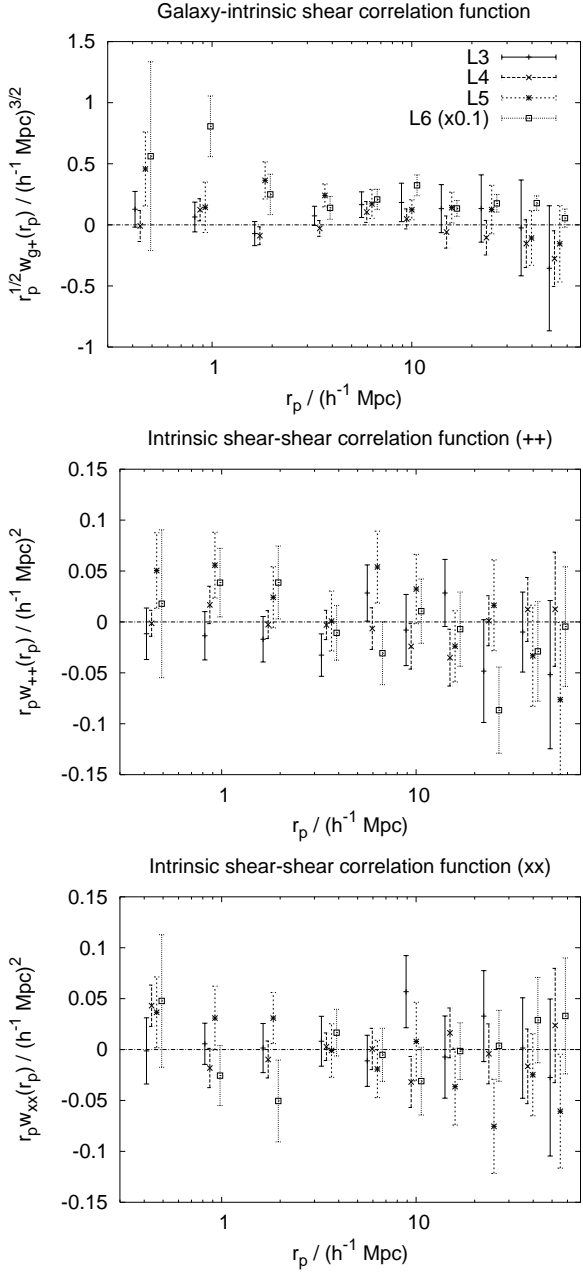}
\caption{\label{fig:mus}The correlation 
functions $w_{g+}(r_p)$, $w_{++}(r_p)$, and $w_{\times\times}(r_p)$ 
obtained from the L3--L6 galaxy samples.  Same as 
Fig.~\ref{fig:rachel} except that Pipeline II was used, and results
are not shown for the full sample.}
\end{figure}

\subsection{Implementation}\label{SS:implementation}

We wrote two pipelines to compute the correlation functions, described 
below.  Pipeline I is based on a tree correlation function code kindly 
provided by Ryan Scranton, and expanded to include quantities with spin 
(e.g. ellipticities), the use of bins in $\Pi$ in addition to $r_p$,
and to make the jackknife subregions have sides that are roughly
equal; the jackknife method is used to obtain error  
estimates.  Pipeline II is a ``brute force'' correlation function code 
designed to be as simple as possible; 
its main purpose is to provide a completely independent check on the much 
more sophisticated Pipeline I.  
Pipeline I was used for the main scientific results of this paper (except 
where indicated), but we have checked that these are not significantly 
altered by using Pipeline II instead (see the result section). 

In order to find pairs of galaxies, Pipeline I uses the SDSSpix 
package.\footnote{URL: \tt 
http://lahmu.phyast.pitt.edu/\~{}scranton/SDSSPix/} To avoid noise in the 
determination of $D-R$ in the $S_+(D-R)/RR$ estimator of the GI 
correlation function, we use 10 random points for each real galaxy in the 
catalog.  The correlation functions are computed over a 120 $h^{-1}$Mpc 
(comoving) range along the line of sight from $\Pi=-60$ $h^{-1}$Mpc to 
$\Pi=+60$ $h^{-1}$Mpc, divided into 30 bins 4 $h^{-1}$Mpc in size, and the 
projected correlation function is computed by ``integration'' (technically 
summation of the correlation function multiplied by $\Delta\Pi$) over 
$\Pi$.  This calculation is done in $N_{bin}=10$ radial bins from 
$0.3<r_p<60$ $h^{-1}$Mpc.  (Note, however, that to avoid calculating pairs 
over excessively large separations, we imposed a cut such that the maximum 
angular separation considered is equal to that at $60$ $h^{-1}$Mpc at 
$z=0.05$, so the relatively small number of lower-redshift pairs do not 
contribute to the result at the largest values of $r_p$.)  Covariance 
matrices are determined using a jackknife with 50 regions.  This number 
was chosen to be large enough to obtain a stable covariance matrix for the 
fits (it must be larger than $N_{bin}^{3/2}$; see Appendix D of 
\citealt{2004MNRAS.353..529H}) but small enough that the size of a 
given jackknife region is larger than the scale on which the correlation 
is to be measured.  Our results for the galaxy-galaxy correlation function 
$\xi(r_p,\Pi)$ and the projected correlation function as a function of 
luminosity match those in \cite{2005ApJ...630....1Z}, a similar analysis 
of SDSS data covering a smaller area of the sky.

Pipeline II is simpler in that for density-shape correlations, 
the $S_+D$ estimator is used instead of $S_+D-S_+R$; and that there is 
only one bin in $\Pi$, $\Delta\Pi = 120$~$h^{-1}$Mpc. Because this
pipeline does not use $D-R$ in any of its estimators, its calculations
were completed with only one random point for every real galaxy
instead of 10.

\subsection{Results}

The correlation functions obtained via Pipeline I are shown in 
Fig.~\ref{fig:rachel}, and via Pipeline II in Fig.~\ref{fig:mus}.  We
note that the difference between the two results is significantly less
than $1\sigma$, indicating that these two independently written
pipelines give the same result. In order to convert $w_{g\times}$ into
$\Delta\gamma_{int}$ in g-g lensing for comparison against
\cite{2004MNRAS.353..529H}, it
is necessary to multiply by -1 (to account for the different
ellipticity convention) and divide by $\Delta\Pi + w_{gg}(r_p)$, where
$\Delta\Pi=120 h^{-1}$Mpc and the projected correlation function can
be estimated from \cite{2005ApJ...630....1Z}, for which the luminosity
subsamples have 
nearly the same selection as in this work. See Appendix A of
\cite{2004MNRAS.353..529H} for more details. Note that the results are 
still not quite appropriate to compare against those in \cite{2004MNRAS.353..529H},
because the $\Delta\gamma_{int}$ so 
obtained from, e.g., our L6 results here would be that expected if one
did g-g lensing with L6 lenses and sources only, instead of L6 lenses
and all sources.  

\subsection{Systematics tests}

\subsubsection{45-degree tests}

The 45-degree rotation of all sources is one of 
the standard tests of systematics. 
We compute both $w_{g\times}$ and $w_{+\times}$ in a manner
analogous to the computation of the signal described in
\S\ref{SS:implementation}, using Pipeline I only.  This computation
was done for all four luminosity bins.  We defer analysis of the
results of these systematics tests to \S\ref{sss:pl}, where we show
consistency of the 45-degree signals with zero.

\subsubsection{Large line-of-sight separation}\label{S:largepi}

Another test that was performed was to compute the signal in the usual
manner, but instead of integrating along the line of sight from $-60$
to $60$ $h^{-1}$Mpc, we integrate in the two ranges satisfying $30 \le
|\Pi| < 90$ $h^{-1}$Mpc.  If any signal is found for $w_{g+}$, $w_{++}$,
or $w_{\times\times}$ when computed in the usual manner, computing it
over large line-of-sight separations with a null result can allow us to
verify that the signal is truly of astrophysical origin rather than
due to some systematic. Results described below (\S\ref{sss:pl})
reveal no evidence of a systematic.

\subsubsection{Jackknife errors}

In this section, we consider the accuracy of the jackknife errorbars,
a crucial part of determining the significance of any
detections and of placing constraints on non-detections.  There are
two competing concerns in determining the number of jackknife regions:
the first is that there should be sufficient regions ($\gg
N_{bin}^{3/2}$, where $N_{bin}=10$) that the
covariance matrix is not too noisy; the second is that the regions
must be large enough that they are truly statistically independent.
If the regions are too small compared to typical pair separations,
then the lack of statistical independence could lead to
underestimation of errorbars, and therefore overestimation of the
significance of detections or overly tight constraints on non-detected
quantities.  

%Our use of 50 jackknife regions, an attempted compromise between these
%concerns, means that the average region is roughly 68 square degrees,
%or a square with sides of length 8.3 degrees.  The largest angular
%separation considered, at the maximum comoving separation of 60
%$h^{-1}$Mpc at $z=0.05$, is 23 degrees.  Consequently, we may be
%concerned that our errorbars will be overly influenced by edge
%effects.  
In order to test whether our errors are overly influenced by edge
effects, we redid the 
comparisons of $w_{g+}(r_p)$ (using the $S_+D/RR$ estimator), $w_{++}(r_p)$, and
$w_{\times\times}(r_p)$ for L3 (which is most prone to edge effects due to its
low redshift) using 25, 50, 75, 100, and 200 jackknife regions.  Our
expectation is that if edge effects are truly significant, then we
should see the size of the errors at large $r_p$ decrease when the number of
jackknife regions is increased; our hope is that the errors will have
converged for our choice of 50 regions.

We also note that jackknife covariance matrices indicate a high
correlation between radial bins on large scales, where the
correlations are higher for the lower luminosity (and therefore lower
redshift) bins since a given transverse separation corresponds to a
larger angular separation.  For example, the correlation coefficient
for $w_{g+}$ between the two outermost radial bins is 0.90, 0.95,
0.73, and 0.45 for L3, L4, L5, and L6 respectively.  All analysis in
the following sections includes the full covariance matrices so that
the correlations are taken into account.

\begin{figure}
\includegraphics[width=3.2in]{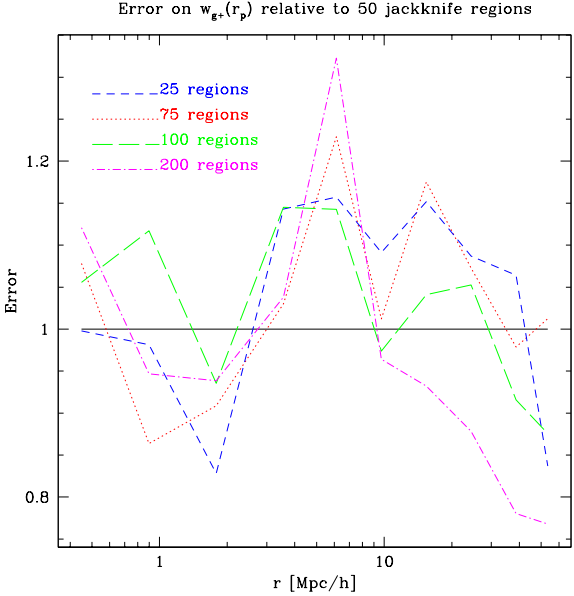}
\caption{\label{fig:comperr} The size of errors of $w_{g+}$ for L3 as a 
function of $r_p$ for the density-shape correlations, relative to the size 
of the errors for 50 jackknife regions.}
\end{figure}
As shown in Fig.~\ref{fig:comperr}, while the errors in $w_{g+}(r_p)$
are clearly noisy 
at the 15 per cent level (which is unsurprising, since 50 is not that
much larger than $10^{3/2} \approx 32$), there do not appear to be
any significant 
edge effects.  While the errors for the largest numbers of samples do
clearly decrease on large scales relative to the errors with smaller
numbers of samples, it is not apparent that there is any statistical
significance to this decrease, and even if it is significant, it is
not a problem for our choice of 50 jackknife regions.  We have also
confirmed that edge effects are not important 
for $w_{++}(r_p)$ and $w_{\times\times}(r_p)$.  We can thus conclude that our
choice of 50 jackknife 
regions should not cause any systematic underestimate of the errors.
We address the question of whether our use of 50 jackknife regions may
lead to overly noisy covariance matrices that affect our results in
\S\ref{sss:pl}. 

\section{Estimates of contamination}\label{S:contam}

In this section we estimate limits on the intrinsic alignment 
contamination for the current and future cosmic shear surveys.  Our
computations here use the vanilla $\Lambda$CDM cosmology with the 
parameters obtained by \citet{2005PhRvD..71j3515S}: $\Omega_b=0.046$, 
$\Omega_m=0.28$, $n_s=0.98$, $\sigma_8=0.9$, and 
$H_0=71.0\,$km$\,$s$^{-1}\,$Mpc$^{-1}$.  Note however that the fractional 
error bars on our intrinsic alignment measurements are much greater than 
those on the cosmological parameters, so reasonable variations in the 
cosmology should not substantially affect our results.  We begin by 
fitting several intrinsic alignment models to the data 
(\S\ref{ss:models}).  We then consider methods of extrapolating these 
models from the SDSS redshift $z\sim 0.1$ to the redshifts $z\sim 1$ of 
interest for cosmic shear (\S\ref{ss:z}).  Then the implications for 
current (\S\ref{ss:current}) and future (\S\ref{ss:future}) cosmic shear 
surveys are discussed.

\subsection{Model fits}
\label{ss:models}

We consider several models for the intrinsic alignments, which can be 
used to estimate the amount of contamination to weak lensing power 
spectra.  The following types of models are considered:

\newcounter{mfit}
\begin{list}{\arabic{mfit}.}{\usecounter{mfit}}
\item {\em Power law fits:} The simplest model is a power-law fit to the 
intrinsic alignment correlation functions $w_{g+}$, $w_{++}$, and 
$w_{\times\times}$.
\item {\em HRH* model:} This model is designed to make our results 
comparable to those of \citet{2004MNRAS.347..895H}, who presented this
modified version of the HRH model originated by
\citet{2000MNRAS.319..649H}.  It fits Eq. \ref{eq:hrhstar} below
to the shape-shape correlation measurements with a free 
amplitude.
\end{list}

In any of these models, once $w_{g+}$, $w_{++}$, and $w_{\times\times}$ 
are determined, we can convert to $P_{g,\tilde\gamma^I}$, 
$P_{\tilde\gamma^I}^{EE}$, and $P_{\tilde\gamma^I}^{BB}$ using the Hankel 
transform relations (Eqs.~\ref{eq:j2}--\ref{eq:jb}).  In the latter two 
cases, the power spectra are immediately useful for intrinsic alignment 
studies.  In the former case, a conversion from $g$ to $\delta$ must be 
applied.  The simplest method here is the linear bias assumption, 
$\delta=g/b_g$, which is valid for some bias $b_g$ on sufficiently large 
scales.   The values of $b_g$ 
\citep{2004ApJ...606..702T, 2005PhRvD..71d3511S} are given in 
Table~\ref{tab:lum}, and we attempt to determine the scales on which
the linear bias assumption is trustworthy using figure 9 in
\cite{2004ApJ...614..533T}, which shows the stochasticity and bias as
a function of scale determined in terms of correlation functions,
i.e. $r = \xi_{gm}/\sqrt{\xi_{gg}\xi_{mm}}$ and $b^2 =
\xi_{gg}/\xi_{mm}$, determined using N-body simulations.  As shown there,
the stochasticity is consistent with 1 for all
scales larger than 400 $h^{-1}$kpc.  The bottom panel indicates that
the bias is approximately linear for scales larger than about 5
$h^{-1}$Mpc, but even at 1 $h^{-1}$Mpc scale, the bias is only about
25 per cent different from its value in the linear regime.

\subsubsection{Power law}
\label{sss:pl}

\begin{figure*}
\includegraphics[width=6.5in]{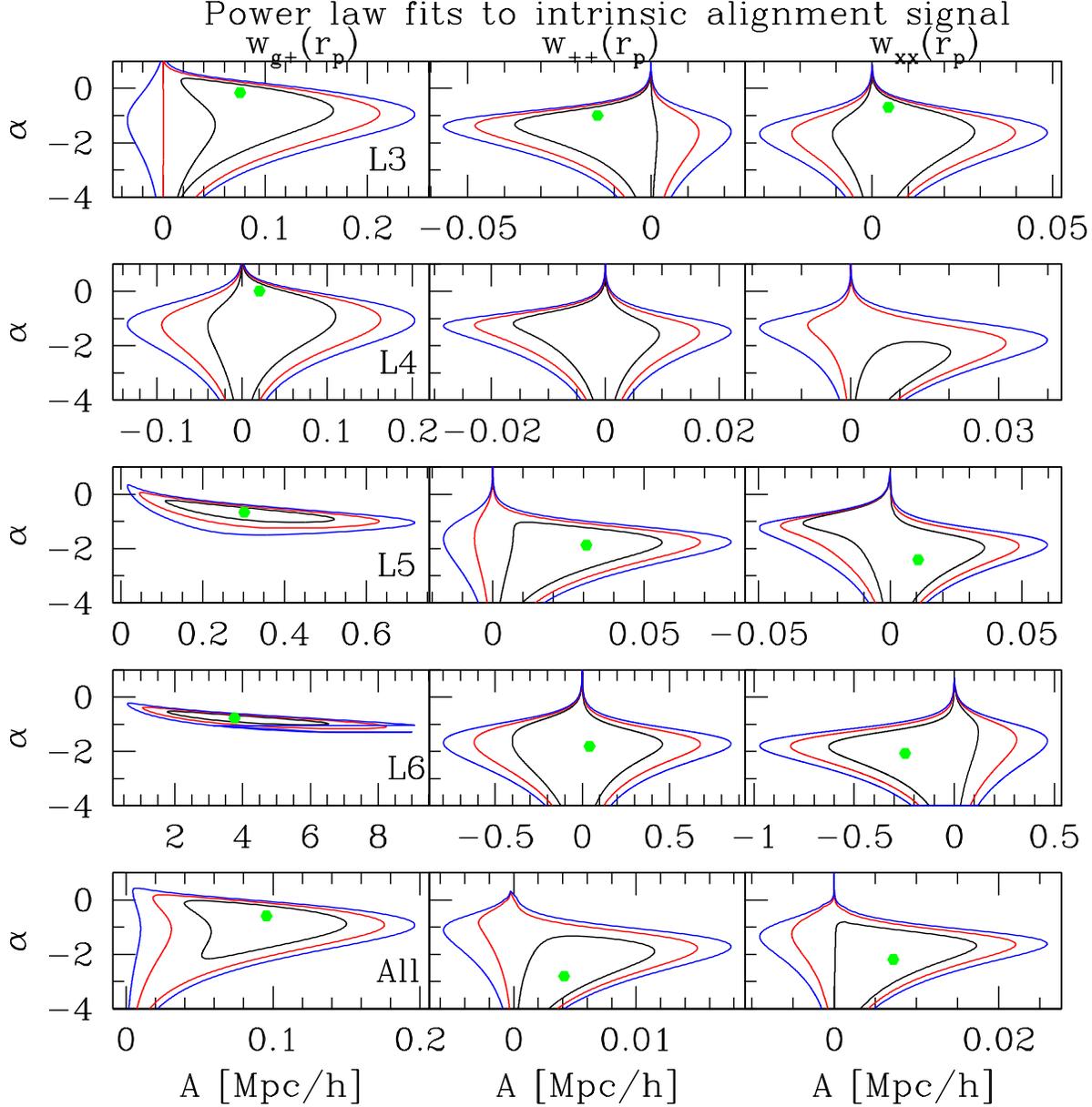}
\caption{\label{fig:powerlaw}Power law fits to the intrinsic alignment 
signal from Pipeline I.  From top to bottom, the rows represent the 
luminosity ranges L3, L4, L5, L6, and the full sample; from left to
right, the columns  
represent the $w_{g+}$, $w_{++}$, and $w_{\times\times}$ correlations 
respectively.  The  dots indicate the $\chi^2$ minimum, and the 
contours represent 75, 95, and 99 per cent confidence regions.  In the L6 
density-shape plot, there was an insufficient number of pairs in the 
innermost bin to establish a reliable jackknife error estimate, so this 
plot is based on only $N=9$ data points.}
\end{figure*}

The power law approach is to fit the signal to the equation:
\begin{equation}
w_{g+}(r_p) = A_{g+}\left(\frac{r_p}{1h^{-1}\,{\rm 
Mpc}}\right)^{\alpha_{g+}},
\label{eq:w-power}
\end{equation}
and similarly for $w_{++}$ and $w_{\times\times}$.  The fits were done by 
computing the $\chi^2$ on a grid in $(A_{g+},\alpha_{g+})$ using the full
jackknife covariance matrix; because this covariance matrix is noisy, the 
usual $\chi^2$ analysis must be modified.  Usually, it is assumed that 
$\Delta\chi^2$ (i.e. the difference in $\chi^2$ between the best-fit 
values of the parameters and the true values) has a $\chi^2$ distribution 
with a number of degrees of freedom equal to the number of parameters 
being fit.  This is no longer the case with a noisy covariance matrix.  In 
Appendix D of 
\citet{2004MNRAS.353..529H}, we examined the distribution of the 
jackknife or bootstrap-derived $\chi^2$ for correlation functions computed 
in $N$ bins and $M$ regions; a similar argument allows one to determine 
$\Delta\chi^2$ for $N$ bins, $M$ regions, and $K$ parameters.  In our 
case, $N=10$, $M=50$, and $K=2$, and the 75th, 95th, and 99th percentiles of 
$\Delta\chi^2$ correspond to $\Delta\chi^2=4.20$, $9.55$, 
and $15.44$ instead of $2.77$, $5.99$, and $9.21$ as derived from the 
standard $\chi^2$ distribution.  Thus, our confidence regions are
larger than those that might be predicted naively using
the standard $\chi^2$ distribution.  The resulting fits and $\Delta\chi^2$ 
contours (which are only strictly accurate for Gaussian error
distributions) are shown in Fig.~\ref{fig:powerlaw}, and the best-fit
parameters are given in Table~\ref{tab:powerlawfits} with 95 per cent
confidence intervals derived using $\Delta\chi^2=6.17$ (the value for one
variable constraints).  We defer further 
discussion of the apparent detection of nonzero $w_{g+}(r_p)$ in L5
and L6 to section~\ref{S:BCG}.  The last row in
Fig.~\ref{fig:powerlaw} shows confidence regions derived from the full
sample of galaxies.  It is clear that the majority of our constraints
come from L4. 

\begin{table}
\caption{\label{tab:powerlawfits}Best-fit parameters for power-law fits
  $A [r_p/(1 \mbox{Mpc}/h)]^{\alpha}$ to the intrinsic alignment
  signal; the 95 per cent confidence intervals shown here may include
  no constraint on $\alpha$ if the amplitude is consistent with zero
  at this level.}
\begin{tabular}{cccc}
\hline\hline
Sample & function & $A$ ($h^{-1}$Mpc) & $\alpha$ \\
\hline
    & $w_{g+}(r_p)$ & $0.082^{+0.106}_{-0.079}$ & $-0.18^{+\infty}_{-\infty}$ \\
L3  & $w_{++}(r_p)$ & $-0.018^{+0.027}_{-0.025}$ & $-1.13^{+\infty}_{-\infty}$ \\
    & $w_{\times\times}(r_p)$ & $0.005^{+0.030}_{-0.022}$  & $-0.68^{+\infty}_{-\infty}$ \\
\hline
    & $w_{g+}(r_p)$ & $0.020^{+0.115}_{-0.085}$ & $0.013^{+\infty}_{-\infty}$ \\
L4  & $w_{++}(r_p)$ & $(-5.7^{+1972}_{-1314})\times 10^{-5}$ & $-5.5^{+\infty}_{-\infty}$ \\
    & $w_{\times\times}(r_p)$ & $(3.8^{+259}_{-3.8})\times 10^{-4}$ & $-7.1^{+5.8}_{-\infty}$ \\
\hline
    & $w_{g+}(r_p)$ & $0.30^{+0.28}_{-0.22}$ & $-0.66^{+0.54}_{-0.46}$ \\
L5  & $w_{++}(r_p)$ & $0.031^{+0.035}_{-0.031}$ & $-1.9^{+1.2}_{-\infty}$ \\
    & $w_{\times\times}(r_p)$ & $0.011^{+0.030}_{-0.029}$ & $-2.4^{+\infty}_{-\infty}$ \\
\hline
    & $w_{g+}(r_p)$ & $3.8^{+3.5}_{-2.2}$ & $-0.77^{+0.29}_{-0.30}$\\
L6  & $w_{++}(r_p)$ & $0.04^{+0.45}_{-0.48}$ & $-1.8^{+\infty}_{-\infty}$\\
    & $w_{\times\times}(r_p)$ & $-0.25^{+1.05}_{-0.49}$ & $-2.1^{+\infty}_{-\infty}$\\
\hline
    & $w_{g+}(r_p)$ & $0.098^{+0.067}_{-0.069}$ & $-0.59^{+0.65}_{-2.30}$ \\
All & $w_{++}(r_p)$ & $(4.3^{+9.3}_{-4.3})\times 10^{-3}$ & $-2.8^{+\infty}_{-\infty}$ \\
    & $w_{\times\times}(r_p)$ & $(7.2^{+13.0}_{-7.2})\times 10^{-3}$ & $-2.1^{+\infty}_{-\infty}$ \\
\hline\hline
\end{tabular}
\end{table}

We also performed the power-law analysis on the 45-degree rotated signals, 
$w_{g\times}(r_p)$ and $w_{+\times}(r_p)$, for each luminosity bin.  Of 
the 8 signals, 7 were consistent with zero at the 68 per cent confidence 
level for $-4\le\alpha\le 1$, and 1 was inconsistent with zero at that level 
but was consistent with zero at the 95 per cent confidence level.  
Consequently, we conclude that this test does not reveal any systematics 
contaminating our analysis.

Finally, we performed the power-law analysis for $w_{g+}(r_p)$ in L5
and L6 (the two bins with apparent detections) with large
line-of-sight pair separations only, as described in
\S\ref{S:largepi}.  The signal computed in this manner is consistent with zero at the 68
per cent confidence level for $-4 \le \alpha \le 1$, which confirms
that the detected signal when computed in
the usual manner is indeed of astrophysical origin.

To demonstrate explicitly the goodness-of-fit of the power-law model,
and the inconsistency with zero for L5, L6, and the full sample,
Table~\ref{T:significance} includes the best-fit parameters, their
$\chi^2$ and $p(<\chi^2)$-value, and $\Delta\chi^2 = \chi^2(A=0) -
\chi^2(\mbox{best fit})$ and its $p(<\Delta\chi^2)$.  (Note that
$p(\chi^2)$ and $p(\Delta\chi^2)$ represent the probability for a
random vector to yield lower values of $\chi^2$ or $\Delta\chi^2$ by
chance, and as such is technically equal to $1-p$, where $p$ is the
probability that the $\chi^2$ will {\it exceed} the given value.  Thus
the $p(<\Delta\chi^2)$ for the $A=0$ case versus for 
the null hypothesis tells us the confidence level at which our results
exclude the null hypothesis.) Because of
the noisiness of jackknife covariance matrices, the $\chi^2$ values do
not actually follow the $\chi^2$ distribution, and the $p(<\chi^2)$
and $p(<\Delta\chi^2)$ values are
therefore computed using a simulation that accounts for this fact (for
more information, see Appendix D of \citealt{2004MNRAS.353..529H}).  Also, to
demonstrate that our results are not overly influenced by 
noisy covariance matrices due to the use of only 50 jackknife regions,
we have repeated the analysis for 100 jackknife regions, and included
the same information for that analysis in Table~\ref{T:significance}.
Results for L6 use the distributions for 9 radial bins; all other bins
use 10. We note that since the use of 100 jackknife regions involves
different weighting of the data, the actual best-fit parameters,
$\Delta\chi^2$ values, and $p$-values are not expected to be precisely
the same as with 50 regions; however, if the fit is good, the best-fit values
should be consistent with each other and the $p$-values for a robust
detection should be small in either case.

There are a few conclusions that we can draw from
Table~\ref{T:significance}.  First, any changes to actual best-fit
parameters due to noisiness of the jackknife covariance matrix (which
would tend to affect the results with 50 jackknife regions more than
with 100 jackknife regions) are
much smaller than the variation allowed within the 95 per cent
confidence intervals on those parameters.  In both cases (50 and 100
jackknife regions), the $\chi^2$
values for all samples except L6 are on the small side, as reflected
in the $p(\Delta\chi^2)$ values tending to all be relatively small.  This may
reflect some overestimation of the errors and therefore
underestimation of detection significance.  The larger $\chi^2$ for L6
may indicate some deviation from power-law behavior, but the actual
value of $\chi^2$ and $p(<\chi^2)$ is not sufficiently large that
using this model is 
unreasonable.  Finally, we consider the $\Delta\chi^2$ values for the
special case of $A=0$ relative to the best-fit $\chi^2$.  As shown,
the results for L3 and L4 are consistent with zero, with the
$\Delta\chi^2$ not being significant.  For both L5 and L6,
high-significance detections were obtained whether we consider the
$\Delta\chi^2$ values using 50 or 100 jackknife regions.  For the full
sample, the detection significance is roughly 99 per cent in both
cases (slightly higher for 50 regions, slightly lower for 100 regions).
\begin{table*}
\caption{\label{T:significance}Best-fit parameters for power-law fits
  $A [r_p/(1 \mbox{Mpc}/h)]^{\alpha}$ to the GI intrinsic alignment
  signal $w_{g+}(r_p)$ (with 95 per cent confidence intervals), the
  $\chi^2$ and $p(<\chi^2)$, and $\Delta\chi^2 = \chi^2(A=0) - 
\chi^2(\mbox{best fit})$ and its $p(<\Delta\chi^2)$.  These numbers are all
  shown for both 50 and 100 jackknife regions.}
\begin{tabular}{ccccc|cc|cccc|cc}
\hline\hline
$\!\!\!$Sample$\!\!\!$ & \multicolumn{6}{c}{50 jackknife regions} & \multicolumn{6}{c}{100 jackknife regions} \\
 & \multicolumn{4}{c}{best-fit} & \multicolumn{2}{c}{($A=0$)-(best fit)}  & \multicolumn{4}{c}{best-fit} & \multicolumn{2}{c}{($A=0$)-(best fit)} \\
 & $\!\!A$ ($h^{-1}$Mpc)$\!\!$ & $\alpha$ & $\chi^2$ & $\!\!p(<\chi^2)\!\!$ & $\Delta\chi^2$ & $\!p(<\Delta\chi^2)$ & $\!\!A$ ($h^{-1}$Mpc)$\!\!$ & $\alpha$ & $\chi^2$ & $\!\!p(\chi^2)\!\!$ & $\Delta\chi^2$ & $\!p(<\Delta\chi^2)$ \\
L3 & $0.082^{+0.106}_{-0.079}$ & $-0.18^{+\infty}_{-\infty}$ & 3.9 & 0.09 & 9.6 & 0.95 & $0.062^{+0.157}_{-0.121}$ & $-0.08^{+\infty}_{-\infty}$ & 4.2 & 0.14 & 3.9 & 0.80 \\
L4 & $0.020^{+0.115}_{-0.085}$ & $0.013^{+\infty}_{-\infty}$ & 5.7 & 0.23 & 2.1 & 0.51 & $0.025^{+0.133}_{-0.102}$ & $-0.008^{+\infty}_{-\infty}$ & 5.1 & 0.22 & 2.1 & 0.58 \\
L5 & $0.30^{+0.28}_{-0.22}$ & $-0.66^{+0.54}_{-0.46}$ & 4.8 & 0.16 & 28.8 & 0.9996 & $0.35^{+0.27}_{-0.26}$ & $-0.73^{+0.60}_{-0.58}$ & 5.7 & 0.28 & 30.8 & $\!\!>0.9999\!$ \\
L6 & $3.8^{+3.5}_{-2.2}$ & $-0.77^{+0.29}_{-0.30}$ & $\!12.6\!$ & 0.83 & 48.2 & $\!\!>0.9999\!\!$ & $3.5^{+4.5}_{-2.4}$ & $-0.72^{+0.70}_{-1.0}$ & $\!11.1\!$ & 0.82 & 54.2 & $\!\!>0.9999$ \\
All & $0.098^{+0.067}_{-0.069}$ & $-0.59^{+0.65}_{-2.30}$ & 5.9 & 0.25 & 18.8 & 0.996 & $0.077^{+0.092}_{-0.077}$ & $-0.35^{+0.65}_{-\infty}$ & 5.5 & 0.26 & 11.0 & $0.987$ \\
\hline\hline
\end{tabular}
\end{table*}

\subsubsection{HRH* model}
\label{sss:hrh}

The HRH* model \citep{2004MNRAS.347..895H} treats the 
intrinsic-intrinsic term according to:
\begin{eqnarray}
w_{++}(r_p) &=& w_{\times\times}(r_p) 
\nonumber \\
&=& \frac{A}{8{\cal R}^2}\int
\left[ 1+ \left(\frac{r}{r_0}\right)^{-\gamma}
\right] \frac{1}{1+(r/B)^2}\, \rmd r_\parallel,
\label{eq:hrhstar}
\end{eqnarray}
where $r=\sqrt{r_\parallel^2+r_p^2}$.  \citet{2004MNRAS.347..895H} fit 
this model with the parameters $\gamma=1.8$, $B=1\,h^{-1}\,$Mpc, and 
$r_0=5.25\,h^{-1}\,$Mpc, leaving $A$ as a free parameter.  Their results 
are shown in Table~\ref{tab:a}; one can see that a detection was achieved 
using SuperCOSMOS.  The original SuperCOSMOS intrinsic alignments analysis 
\citep{2002MNRAS.333..501B} used a median redshift of $z_{med}=0.1$ for 
their sample with limiting magnitude $b_J=20.5$, and this number was also 
used in the re-analysis by \citet{2004MNRAS.347..895H}.  The source of
the quoted $z_{med}$ is a parametric model from
\citet{1993MNRAS.265..145B} using APM data, which was obtained using
fits to a particular form of the redshift distribution including a
parameter $z_{med}(b_J)$.  It is possible that the weak dependence of
$z_{med}$ on $b_J$ for bright galaxies assumed by the fitting formula
may have unduly influenced the 
fits, causing $z_{med}$ to have been underestimated at intermediate
magnitudes.   

However, more  
recent data from the 2 degree Field (2dF) galaxy survey has enabled a 
direct measurement of the $b_J$ band luminosity function and $k+e$-corrections
\citep{2002MNRAS.336..907N}, from which we find $z_{med}\sim 0.16$ for this 
limiting magnitude.  In support of this calculation, we note that from
the 2dF catalog itself (which has a limiting magnitude of
$b_J=19.45$), $z_{med}$ for $b_J<18.5$ is $0.082$, and for $b_J<19.45$
is $0.11$.  If we extrapolate the $b_J<19.45$ number to the
SuperCOSMOS flux limit using the
Euclidean method, $z_{med}\sim 10^{0.2 b_{J,lim}}$ (valid for $z \ll
1$) , we obtain 
$z_{med} = 0.174$.  If we determine an exponential $z_{med} \sim
10^{a\, b_{J,lim}}$ and extrapolate that to $b_{J,lim}=20.5$, we obtain
  $z_{med}=0.144$.  Both of these extrapolations merely serve as
  sanity checks of our result from the luminosity function
  calculation, and the fact that $z_{med}$ for $b_{J,lim}=19.45$ is
  already larger than 0.10 in the 2dF data  further supports our
  assertion that the 
  SuperCOSMOS calculations significantly underestimated $z_{med}$.
  The large underestimate of the median redshift for  
the SuperCOSMOS data increases the theoretical predictions for the
observed signal,  
and hence decreases the best-fit value of $A$; we therefore conclude that 
\citet{2004MNRAS.347..895H} underestimated both the value of $A$ and its 
error bar.  We have re-computed the theoretical predictions for the 
ellipticity variance at $z_{med}=0.16$ and found the correction to be a 
factor of 3.0--3.4 over the range of angular scales considered by 
SuperCOSMOS (25--93 arc minutes).  The value shown in the table has 
therefore had a re-scaling factor of 3.2 applied (using 3.0 or 3.4 instead 
changes the results by $<0.2\sigma$).  Note that the change in the 
redshift distribution does not affect the statistical significance of the 
SuperCOSMOS detection, although there are potential sources of
systematic error in this data set that could lead to a spurious signal.

The constraint on $A$ from the full SDSS sample is slightly tighter than 
that from SuperCOSMOS, but due to its lower central value does not 
provide a detection.   Combining the two results is dangerous because the 
southern stripes of SDSS overlap with the SuperCOSMOS survey area and 
hence the two results are not truly independent.  
The results in Table~\ref{tab:a} suggest that the SDSS and SuperCOSMOS
constraints on  
$A$ are mutually consistent, with any discrepancy below the
$1\sigma$ level and thus not very statistically significant,
particularly in light of the large systematic uncertainty due to the
uncertainty in median redshift.  Indeed, this comparison highlights
the need for accurate redshift distributions in order to properly
constrain the amplitude of intrinsic alignment correlation functions.

\begin{table}
\caption{\label{tab:a}The intrinsic alignment amplitude $A$ in 
Eq.~(\ref{eq:hrhstar}), as determined in this paper and by 
\citet{2004MNRAS.347..895H}.  All results are 95 per cent confidence; the 
error bars from this work are determined using the jackknife confidence 
intervals as described in \S\ref{sss:pl}.  The SuperCOSMOS and COMBO-17 
ellipticity data are described in \citet{2000MNRAS.318..625B} and 
\citet{2003MNRAS.341..100B}, respectively.  The analysis of SuperCOSMOS 
by \citet{2004MNRAS.347..895H} gave a value of $A=(0.9\pm 
0.5)\times 10^{-3}$; we have re-scaled this result by a factor of 3.2 to 
account for the more recent determination of the SuperCOSMOS redshift 
distribution.  See \S\ref{sss:hrh} for details.}
\begin{tabular}{lrl}
\hline\hline
Sample & $A/10^{-3}$ & Reference \\
\hline
SDSS L3 & $-1.0\pm 6.0$ & This work \\
SDSS L4 &  $1.7\pm 4.3$ & This work \\
SDSS L5 &  $5.8\pm 7.4$ & This work \\
SDSS L6 &  $-43\pm 122$ & This work \\
\hline
SDSS L3--L6 & $1.8\pm 2.3$ & This work \\
\hline
COMBO-17 & $<5.4$ & \citet{2004MNRAS.347..895H} \\
\hline
SuperCOSMOS & $2.9\pm 1.6$ & \citet{2004MNRAS.347..895H} \\
%\\ & & this work (see \S\ref{sss:hrh}) \\
\hline\hline
\end{tabular}
\end{table}

\subsection{Redshift evolution}
\label{ss:z}

In order to estimate the amount of intrinsic alignment contamination in 
cosmic shear surveys, one must extrapolate from the SDSS redshift range 
$z\sim 0.05$--$0.2$ to the redshifts of the sources used in these surveys, 
$z\sim 0.3$--$2$.  We attempt several types of extrapolation here.  Note 
that in each case the extrapolation becomes successively more dangerous as 
we go to higher redshift: our results for the lowest-redshift surveys such 
as CTIO are probably very good, and any attempt to make statements at 
redshifts $z\approx 2$ are probably close to meaningless.

Extrapolation in redshift involves three major issues.  One is that the 
intrinsic ellipticities of galaxies may change with time due to (for 
example) mergers.  A second issue is that galaxies move, and thus the 
intrinsic alignment power spectra can change even if the orientation of 
each galaxy remains fixed.  Finally, we have established that the 
intrinsic alignment signal depends on the sample of galaxies considered, 
and it is not clear which sample of galaxies observed at $z=0.1$ is most 
similar to (or is evolved from) a particular sample of galaxies at
higher redshift.  Each of these issues must be addressed in the
context of both the density-shape and shape-shape correlations. 

In the case of the shape-shape correlation, the argument in 
\citet{2004PhRvD..70f3526H} would suggest that in the linear regime, 
$w_{++}(r_p)$ and $w_{\times\times}(r_p)$ would be redshift-independent 
(again assuming the intrinsic ellipticities of individual galaxies do not 
change).  Of course, most of the constraints from lensing, and most of the 
observational data presented here, are from nonlinear scales where the 
galaxy correlation function $\xi_{gg}(r_p,\rpi)\ge 1$.  In the nonlinear 
regime, \citet{2004MNRAS.347..895H} argued that the correlation function 
of the {\em unweighted} intrinsic shear $\bgamma^I$ would vary slowly with 
redshift, while the {\em density-weighted} correlation functions 
$w_{++}(r_p)$ and $w_{\times\times}(r_p)$ would increase at 
low $z$ due to the growth of perturbations $\delta_g$.  In particular, if 
one examines the strongly nonlinear regime $\xi_{gg}\gg 1$, and assumes 
stable clustering with $\xi_{gg}\propto (1+z)^{-3+\gamma}$ ($\gamma\approx 
1.8$ is the slope of the galaxy correlation function), then it follows 
that both $w_{++}(r_p)$ and $w_{\times\times}(r_p)$ should scale as 
$(1+z)^{-3+\gamma}$.

However, if one takes seriously the concept of stable clustering, then the 
unweighted intrinsic shear correlation function, 
$\xi_{\bgamma^I\bgamma^I}$, should be constant when measured in physical 
rather than comoving separation.  In this case, the correlation functions 
of $\tilde\bgamma^I$ at redshift $z$ are
\begin{equation}
\xi_{++}(r_p,r_\parallel;z) = (1+z)^{-3}\xi_{++}\left(
\frac{r_p}{1+z},\frac{r_\parallel}{1+z};0\right).
\end{equation}
Integration over $r_\parallel$ then gives
\begin{equation}
w_{++}(r_p;z) = (1+z)^{-2}w_{++}\left(\frac{r_p}{1+z};0\right),
\label{eq:sc-ss}
\end{equation}
and similarly for $w_{\times\times}(r_p)$.

In the case of the density-shape correlation, the simplest prescription 
was provided by \citet{2004PhRvD..70f3526H}, who argued that if the 
intrinsic ellipticities of individual galaxies did not change, then on 
linear scales $w_{\delta+}(r_p)$ would scale as the growth factor.  On the 
other hand, on nonlinear scales, stable clustering would suggest
\begin{equation}
w_{\delta+}(r_p;z) = (1+z)^{-2}w_{\delta+}\left(\frac{r_p}{1+z};0\right),
\label{eq:sc-ds}
\end{equation}
in analogy to Eq.~(\ref{eq:sc-ss}).  The analyses below consider both the 
linear evolution and the stable clustering methods for extrapolating the 
signal to high redshift.  We do not consider other methods of 
extrapolation since these would likely be much more complicated and it 
is not clear that they would have higher fidelity.

\subsection{Current surveys}
\label{ss:current}

In the past several years, a number of cosmic shear surveys have reported 
results for $\sigma_8$, which is the cosmological parameter most readily 
accessible to cosmic shear.  It is apparent that there is some 
disagreement between the determinations (see, e.g., Table 1 of 
\citealt{2005astro.ph..6112H}).  While the differences may arise due to 
shear calibration biases of up to 15 per cent between the PSF correction 
methods used (see, e.g., \citealt{2003MNRAS.343..459H} and 
\citealt{2005astro.ph..6112H}) or uncertainties in $z_m$, there is also 
the possibility that the samples used for each of these studies were 
contaminated by intrinsic alignments in very different ways.  
Unfortunately, the determination of how our results apply to these samples 
is highly nontrivial.  We have selected our galaxies in the $r$ band at 
$z\sim 0.1$, but the cosmic shear selection at $z\sim 0.6$--$1.0$ is 
usually in the observed-frame red/far-red bands (similar to SDSS $r$ or 
$i$), which is frequently in the rest-frame ultraviolet.  This makes it 
very difficult to determine whether the samples in those tables correspond 
to the lower-contamination $L<L_*$ samples, or the higher-contamination 
$L>L_*$ samples.

In principle, one possible method of determining the correspondence
between samples would be to use a simple one-parameter family of
templates from the stellar population synthesis code described in
\cite{2003MNRAS.344.1000B} with exponential star-formation rates.
Even this simple model can be shown to accurately reproduce the colors
of SDSS main spectroscopic sample galaxies over for $0<z<0.25$.  One
could then find the regions in the 
color-color and color-magnitude diagrams corresponding to each of the
luminosity bins, use the models to evolve them in redshift, and see what
they looked like at $z\sim 0.6$--$1.0$.  However, such a procedure
entails enough systematic uncertainty that it is unclear whether the
results would be trustworthy.  Hence, we defer a detailed discussion of how the
detected GI alignments contaminate the results of current surveys to
future work.

It is interesting to note that on small scales 
the GI effect on the observed angular shear 
correlation function may have been detected in the 
Red-sequence Cluster Survey (RCS).  \citet{2004ApJ...606...67H} observed 
an anticorrelation between the shapes of bright ($19<R_C<21$) and faint 
($21.5<R_C<24$) galaxies.  This anticorrelation is not consistent with 
either cosmic shear or estimates of the systematic errors in RCS, and 
\citet{2004ApJ...606...67H} attributed it to the alignment of the bright 
galaxies with the ellipticities of their dark matter haloes (which lens 
the more distant, faint galaxies).\footnote{\citet{2005astro.ph..7108M} 
did not find such a correlation in the SDSS, but the samples of galaxies 
used were very different, and both \citet{2005astro.ph..7108M} and this 
work suggest that these correlations are strongly dependent on the sample 
of galaxies considered.}  The halo ellipticity will tend to lead to a
stronger lensing signal along the lens major axis, i.e. stronger
source tangential ellipticity along this axis, and therefore an
anticorrelation of lens and source ellipticities.  This halo
ellipticity effect is a  
special case of the density-intrinsic shape correlation, although 
the physical explanation in terms of alignment of galaxy light with 
halo mass distribution suggests it would be limited to scales 
below the virial radius of the halos.  On larger scales, such an
anticorrelation can be interpreted as ellipticity of local large-scale
structure rather than of individual halos.   
Of course, the galaxy samples in 
\citet{2004ApJ...606...67H} were selected by putting most of the 
``lenses'' at lower redshift than the ``sources,'' which 
\citet{2004PhRvD..70f3526H} argued maximizes the GI effect.  
Nevertheless, precisely this type of separation of source screens has been 
proposed for cosmic shear tomography studies.  The apparent detection of 
halo ellipticity also suggests that in some cases, the GI effect can 
exceed 100 per cent of the GG effect and result in negative shear 
correlation functions.  Of course, such regimes must be avoided for cosmic 
shear studies.

\subsection{Future surveys}
\label{ss:future}

\begin{figure}
\includegraphics[angle=-90,width=3.2in]{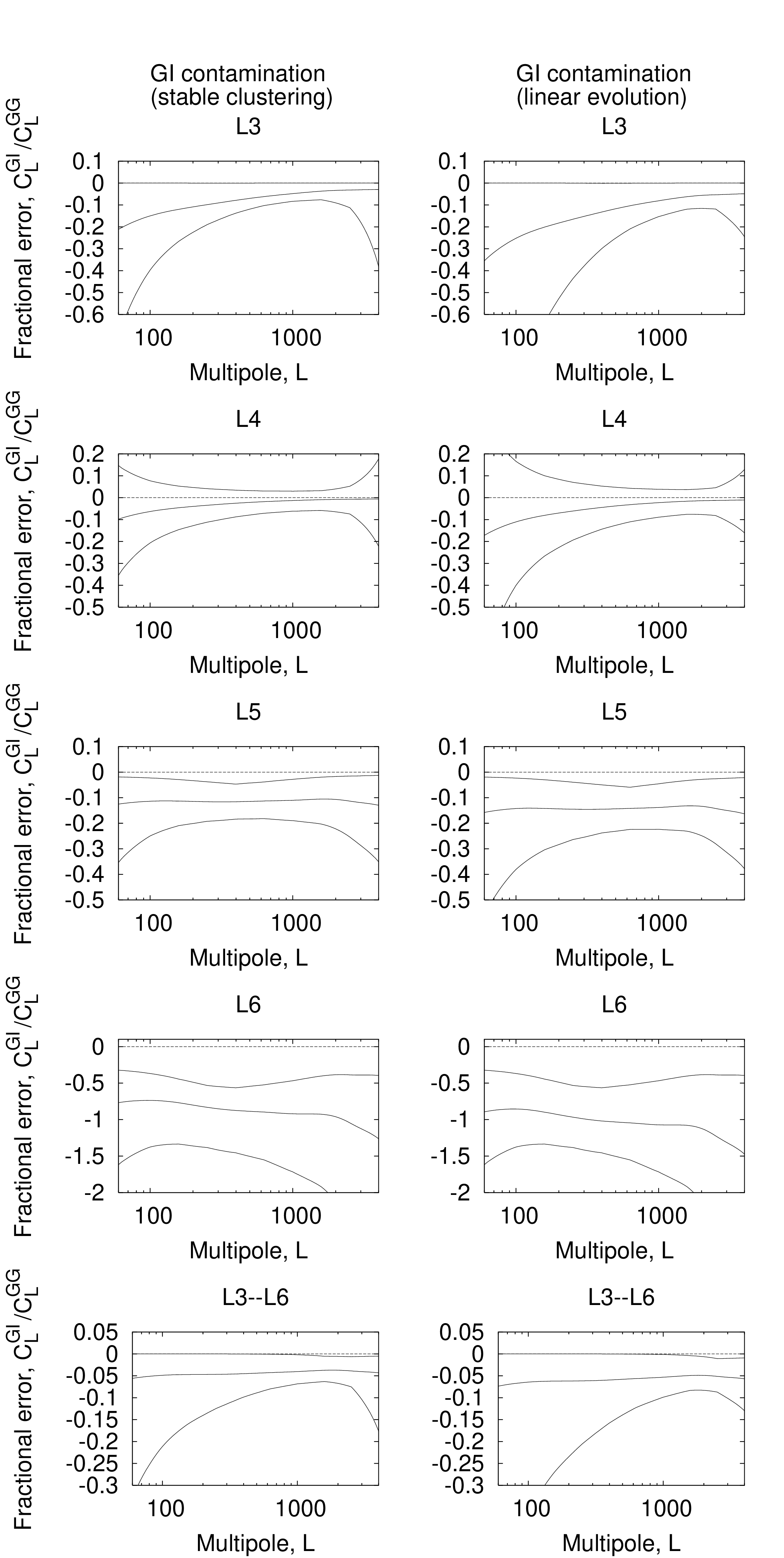}
\caption{\label{fig:gicontam}The allowed range of GI contamination for 
each luminosity subsample.  The power-law fits are used, with the left 
column showing the results for the stable clustering assumption and the 
right column showing the results for linear evolution as argued by 
\citet{2004PhRvD..70f3526H}.  The bottom and top curves show the 95 per 
cent confidence region assuming a power law intrinsic alignment model 
with index $-3<\alpha_{g+}<+1$.  The center curve shows the contamination 
predicted by the best-fit parameters in Table~\ref{tab:powerlawfits}.  The 
median source redshift assumed is $z_{med}=1.0$.  
[The constraints on $\alpha_{g+}$ are imposed because for $\alpha_{g+}\ge 
+1$ or $\alpha_{g+}\le -4$, the Hankel transform defining 
$P_{\delta,\tilde\gamma^I}(k)$ becomes ill-defined.  A cutoff value 
greater than $-4$ was chosen because otherwise the correlations at very 
small scales dominates the power spectrum.  Note that in the cases of L5 
and L6 where we have a detection, $\alpha_{g+}$ is constrained to lie 
within the range given here at $>99$ per cent confidence.]}
\end{figure}

\begin{figure}
\includegraphics[angle=-90,width=3.2in]{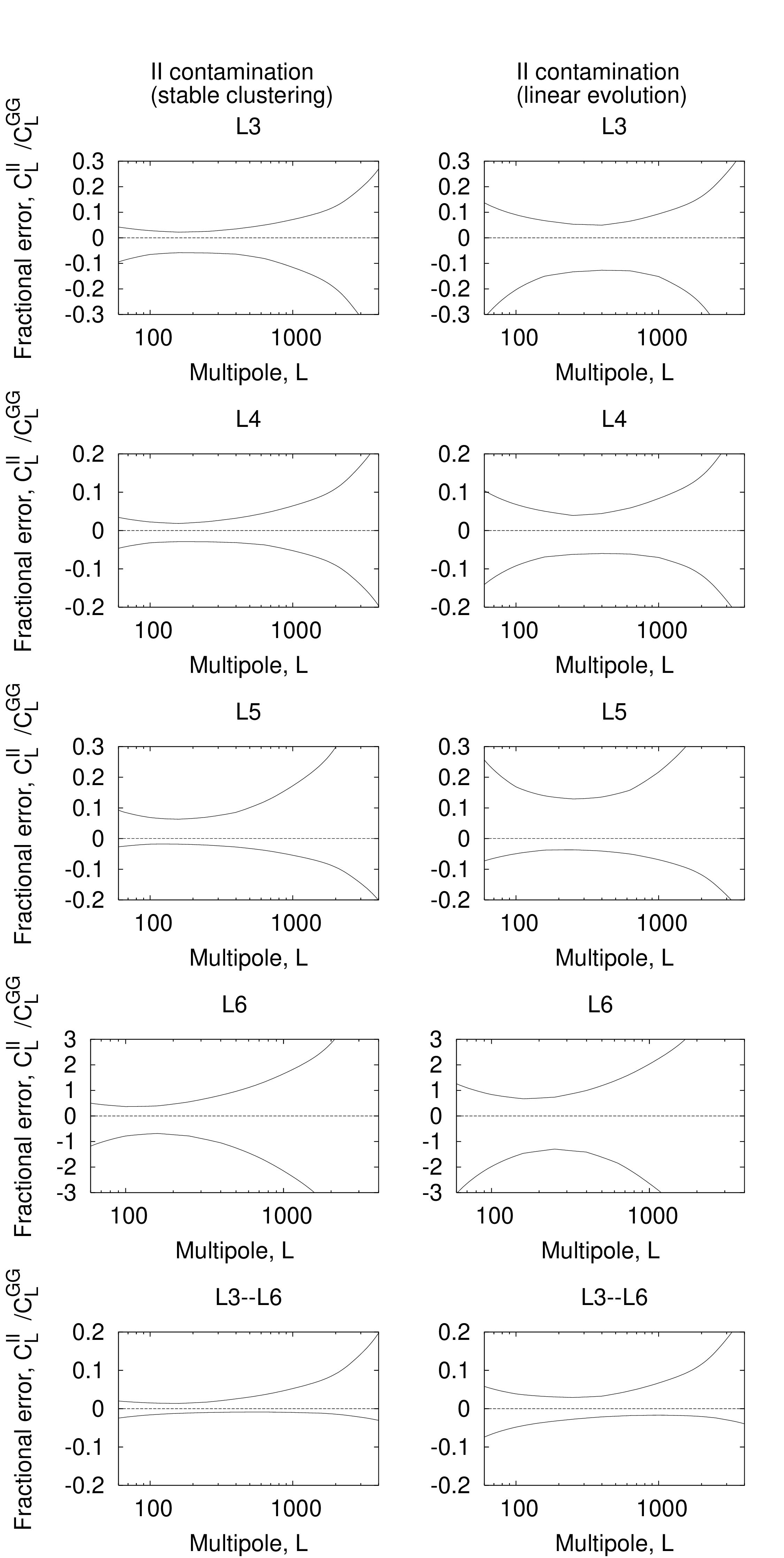}
\caption{\label{fig:iicontam}The allowed range of II contamination for
each luminosity subsample.  The power-law fits are used, with the left
column showing the results for the stable clustering assumption and the
right column showing the results for linear evolution as argued by
\citet{2004PhRvD..70f3526H}.  The bottom and top curves show the 95 per
cent confidence region assuming a power law intrinsic alignment model
with index $-1.5<\alpha<0$. The median source redshift assumed is 
$z_{med}=1.0$.
[The constraints on $\alpha$ are imposed because for $\alpha_{++}\ge
+0$ or $\alpha_{++}\le -2$, the Hankel transform defining
$P_{\delta,\tilde\gamma^I}(k)$ becomes ill-defined.  A cutoff value
greater than $-2$ was chosen because otherwise the correlations at very
small scales dominates the power spectrum.  We do not detect II in any of 
these cases, so the actual power law slope cannot be determined from the 
data.]}
\end{figure}

The GI contamination estimated using the power-law fits is shown in 
Fig.~\ref{fig:gicontam} and the II contamination in 
Fig.~\ref{fig:iicontam} for a cosmic shear survey with a median redshift 
of $z_{med}=1.0$.  The full sample (L3--L6) GI results should be treated 
with caution because they are obtained by averaging the various galaxy 
samples whose GI contamination amplitudes are not consistent with each 
other.  The average is thus appropriate to the SDSS spectroscopic
sample, or other samples at the same
redshift range with the same selection criteria, but may not be a 
good guide otherwise.  In particular, when considering GI 
contamination for a future survey, one must determine which of the 
luminosity subsamples is most similar to the sources to be used in that 
survey.

\begin{figure}
\includegraphics[angle=-90,width=3.2in]{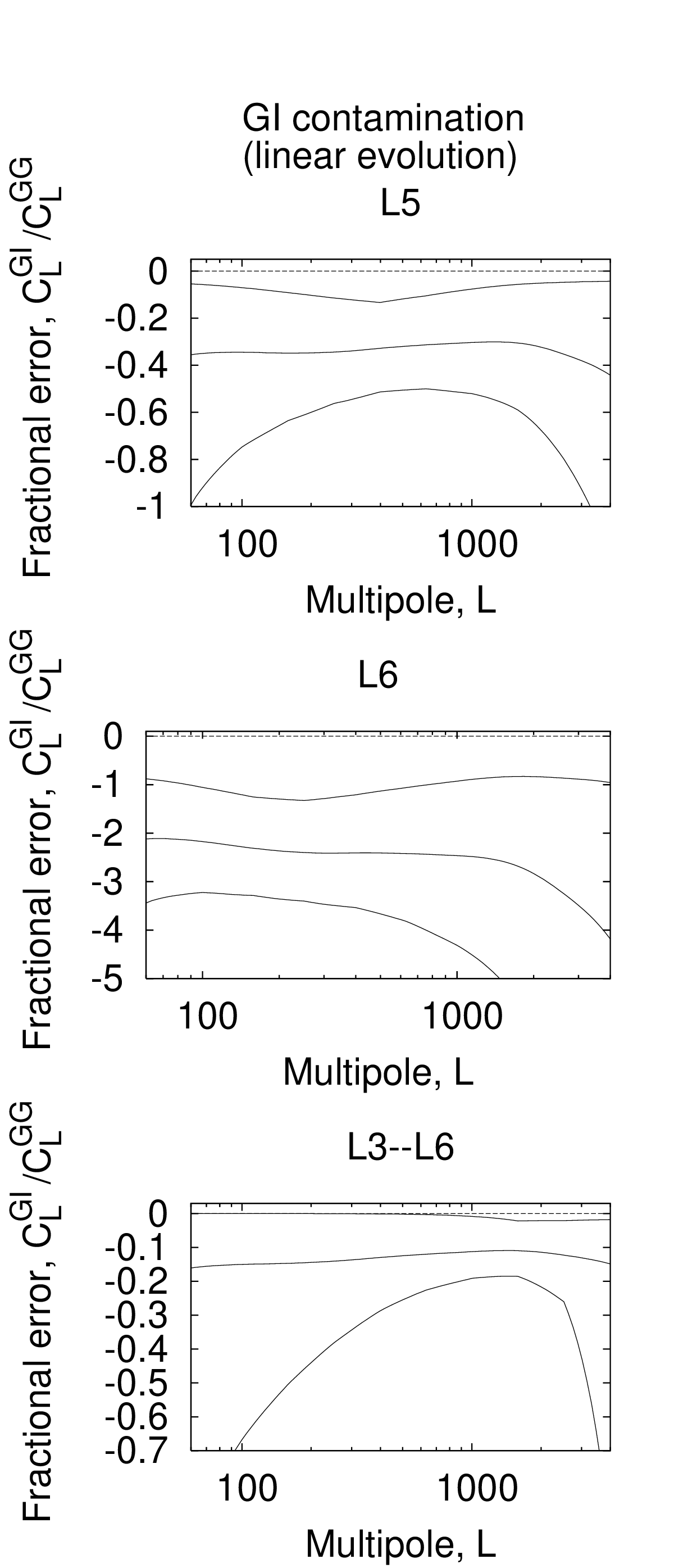}
\caption{\label{fig:gilr}The allowed range of GI contamination for 
L5, L6, and the full sample determined using the power-law fits and
linear evolution for $z_{med}=0.6$.  The bottom and top curves show the 95 per 
cent confidence region assuming a power law intrinsic alignment model 
with index $-3<\alpha_{g+}<+1$.  The center curve shows the contamination 
predicted by the best-fit parameters in Table~\ref{tab:powerlawfits}.}
\end{figure}

Fig.~\ref{fig:gilr} shows the estimated fractional GI contamination
for a survey with $z_{med}=0.6$ (determined using linear evolution,
though both methods give nearly identical results for these low
redshifts) only for those samples that had a
statistically significant determination of signal.  A comparison
between Figs.~\ref{fig:gicontam} and~\ref{fig:gilr} shows that the
contamination is more severe at lower redshifts, as expected.

We note that, as discussed in \S\ref{ss:models}, these estimates of
contamination are not reliable on small scales due to the breakdown of
the linear bias assumption.  At $<1$ $h^{-1}$Mpc, or $L>\sim 1000$ at
$z_{med}=1$, we may be overestimating the bias by approximately 30 per
cent, and consequently underestimating the contamination by that amount.

\subsection{Cosmic shear tomography}

Thus far we have only discussed the effects of intrinsic alignments on
cosmic shear autocorrelation studies.  In this section, we also
consider the effects on cosmic shear tomography, the cross-correlation
of pairs widely separately in redshift, either via down-weighting
pairs close in photometric redshift, or by explicitly separating into
multiple samples based on photometric redshift or apparent magnitude such
that the samples lie in distinct regions in redshift.  In principle, while the
cross-correlation should still give a GG signal due to lensing
by structures between the observer and the lower-redshift source
sample, any intrinsic alignment II contamination will be highly
suppressed.  However, as pointed out in \cite{2004PhRvD..70f3526H},
the GI contamination may still be significant when selecting pairs
separated in redshift space, because the galaxy at lower redshift can
be aligned with the same tidal field that lenses the 
higher-redshift galaxy of the pair.  In light of our robust detection
of the GI contamination using pairs at the same redshift, more
investigation into the effect of this contamination on cosmic shear
tomography is warranted; this investigation is beyond the scope of
this paper. Some of these issues have been investigated by \cite{2005A&A...441...47K}. 

\section{Intrinsic alignments and BCGs}\label{S:BCG}

An early study of alignments of cluster galaxy ellipticities with the 
cluster major axis, \cite{1982A&A...107..338B}, found a tendency for 
brightest cluster galaxies (BCGs) to align with the cluster 
ellipticity using 44 Abell clusters with $z<0.1$. Later studies confirmed 
this ``Binggeli effect,'' \cite{1999ApJ...519...22F} with poor clusters, 
\cite{2000ApJ...543L..27W} in 3 dimensions for the Virgo cluster, and
\cite{2002ASPC..268..395K} using $\sim 300$ clusters in  
SDSS data over a large range in redshift, $0.04<z<0.5$.  This effect has 
been explained in terms of anisotropic infall along filaments 
\citep{1989ApJ...347..610W}, an explanation that is supported by numerous 
$N$-body simulations (e.g. \citealt{1998ApJ...502..141D}).  [The
related ``Holmberg effect'' \citep{1969ArAst...5..305H}, the
correlation of satellite positions with the major axis of the primary
(not in clusters), has also been studied fairly extensively observationally
\citep{1975AJ.....80..477H,1978AJ.....83..135V,1982MNRAS.198..605M, 
1997ApJ...478L..53Z, 2004MNRAS.348.1236S, 2005ApJ...628L.101B} and
with simulations
\citep{2004ApJ...613L..41N,2005astro.ph..6547A,2005A&A...437..383K,2005MNRAS.363..146L,2005ApJ...629..219Z},
and is another possible explanation, though unlike for clusters the
results -- magnitude and even sign of the effect -- have been
conflicting.]  This effect may be used to explain the  
detection of nonzero $w_{g+}(r_p)$ in L5 and L6: if some of these 
bright galaxies are BCGs of clusters, then we expect to find an overdensity of 
galaxies along their major axis, leading to positive $w_{g+}$.  In
support of this hypothesis, we note that the majority of these bright
galaxies are red, and many pass the color cuts to be included in the SDSS
spectroscopic LRG sample; it is well known that BCGs tend to be red
galaxies.

Naturally, BCG alignments can only explain a detection on small
($<\sim 2$ $h^{-1}$Mpc) scales.  However, when combined with the
result \citep{2005ApJ...618....1H} that clusters themselves are
aligned with each other for 
separations up to 100 $h^{-1}$Mpc, we can also explain the larger
scale alignments, since it means that a BCG of a given cluster also
tends to point preferentially towards overdensities of galaxies on
larger scales and not just within its own cluster.  We also note the
result shown there (from $N$-body simulations) that the alignment
increases with increasing redshift, indicating that if our explanation
of the effect is correct, then it may be worse for higher-redshift
surveys than what has been detected here.

If indeed the major source of density-shape correlation is BCG alignment, 
it may be advantageous for cosmic shear surveys to reject BCGs, 
particularly in the lower-redshift bins where their intrinsic alignments 
can contaminate the cross-correlation tomography signal.  A variety of 
algorithms exist to identify clusters and their BCGs (e.g.
\citealt{2000AJ....120.2148G, 2002AJ....123.1807G, 2003ApJS..148..243B}), 
and it is beyond the scope of this paper to examine how much the 
density-shape correlation can be suppressed by rejecting BCGs selected by 
each of these methods.

\citet{2005ApJ...627L..21P} use X-ray selected clusters with SDSS
photometric and spectroscopic imaging data to find a tendency for
radial alignment of cluster galaxies, not limited to BCGs.  To explain
their findings, they propose a parametric resonance phenomenon that
causes a tendency for many cluster galaxies to exhibit radial
alignment relative to the cluster center.  Their results suggest that
removing BCGs from the sample may not be sufficient to eliminate GI
contamination.  Further investigation of our results, including a
search for intrinsic alignments that distinguishes between satellites
in clusters and the BCGs themselves, is necessary to understand the
degree to which the effect they observe is important for cosmic shear
surveys. 

We also note that the effect seen here is related to
the effect seen by correlating positions of a fainter set of galaxies with the
position angles of lens galaxies in \cite{2005astro.ph..7108M}.  There, 
as shown in Figure 6 and the accompanying text, a statistically significant
tendency for the fainter galaxies to be positioned along the major
axis of the central galaxy was found for L5 and L6.

\section{Conclusions}\label{S:conclusions}

We have used the SDSS main spectroscopic sample to
search for II and GI correlations by computing projected correlation
functions $w_{g+}(r_p)$, $w_{++}(r_p)$, and $w_{\times\times}(r_p)$
over the range of pair separations $0.3 < r_p < 60$ $h^{-1}$Mpc.  We
have two main results coming out of these calculations.

The first result is a constraint on the II correlation, expressed in
terms of several models in \S\ref{ss:models} and in terms of projected
contamination for higher-redshift surveys in \S\ref{ss:future}.  As
shown, we have no detection of II correlations either in 
luminosity bins or using the full sample. The limits
rule out a significant contamination of shear signal by II, although 
contamination at the 10 per cent level for lensing surveys at $z_s
\sim 1$ is still a possibility.  Thus, the II correlation may still be
a concern for future cosmic shear surveys with expected statistical
errors below the 1 per cent level, though it can be minimized by
cross-correlating source galaxies in different redshift slices.

The second result is a detection of GI correlations for $L > L_*$ galaxies 
(L5 and L6), as well as for the overall sample, with nonzero amplitude of 
the correlation function $w_{g+}(r_p)$ for the power-law subscript $-4 \le 
\alpha \le 1$, at the $>99$ per cent confidence level.  
We have no
detection of GI correlations for $L\le L_*$ galaxies (L3 and L4), and
have placed constraints on them for the first time.
The agreement of 
our results from two independent pipelines to calculate the correlation 
functions confirms the robustness of our fundings.  As for the II 
correlations, we have used several models of extrapolating these GI 
results to higher-redshift cosmic shear autocorrelation power spectra in 
\S\ref{ss:future} to predict GI contamination at $z_s=1$ and $L=100-1000$ of 
5--25 per cent (L5), 50--150 per cent (L6), or 0--15 per cent (full 
sample). These GI correlations are positive in sign, which is the opposite of 
the GG lensing induced signal, so the interference is destructive
\citep{2004PhRvD..70f3526H}.  As a result the amplitude of fluctuations 
is systematically underestimated in current surveys where this effect
has been ignored. The estimated error in the linear amplitude
$\sigma_8$ from this effect is uncertain,  but estimates presented
here suggest current surveys underestimate it by 0-20 per cent for
$z_s\sim 1$ and possibly up to 30 per cent for shallower surveys with
$z_s \sim 0.5$.  

These results leave open a number of questions for future work.
First, is there some way to minimize the contamination due to the GI
correlations, which cannot be eliminated by the scheme proposed for
eliminating II correlations via cross-correlation of sources at
different redshifts?  We have proposed BCG alignments with cluster
ellipticities as a possible explanation for the correlation, so
eliminating BCGs from the sample used for computing cosmic shear
should be investigated as a means of reducing this alignment, though
as suggested by \citet{2005ApJ...627L..21P}, this step alone may not
be sufficient.  Second, 
what are the implications of the GI detection for current cosmic
shear studies, and could they explain the conflicting values of
$\sigma_8$ (which do have other possible explanations, such as shear
calibration bias or uncertainty in redshift distributions)?  In order
to investigate this question thoroughly, it will be necessary to
determine the correspondence between the samples used for this work
and for the current cosmic shear studies.  Finally, it is clear that a 
systematic error at a 10 per cent level cannot be tolerated in future surveys 
such as LSST, PanStarrs or JDEM, which require a sub-percent precision 
to achieve the stated goals. 
More detailed implications for future work, including cosmic shear tomography, which
is susceptible to contamination from the GI detection, will be addressed
in a following work, \cite{IshakHirataMandelbaum2005}.

In summary, our detection of the GI contamination of the cosmic shear
power spectrum should
serve as a useful starting point for further investigation into this
effect, which is one of the main theoretical uncertainties in
understanding current cosmic shear analyses, and may also lead to
methods that can help reduce this contamination to the lowest level
possible when interpreting data from future surveys.

\section*{Acknowledgments}

RM is supported by an NSF Graduate Research Fellowship.
CH is supported in part by NSF PHY-0503584 and by a grant-in-aid from
the W. M. Keck Foundation.  MI acknowledges the support of the Natural Sciences
and Engineering Research Council of Canada (NSERC) and
NASA Theory Award NNG04GK55G.

We wish to thank Ryan Scranton for the use of his correlation function 
code.  We thank David Spergel for pointing out some relevant
references, Catherine Heymans for useful discussion of results, and
Michael Brown for clarification regarding the SuperCOSMOS dataset.
Finally, we thank the referee for useful suggestions that help
strengthen the conclusions of the paper.

Funding for the creation and distribution of the SDSS Archive has been 
provided by the Alfred P. Sloan Foundation, the Participating 
Institutions, the National Aeronautics and Space Administration, the 
National Science Foundation, the U.S. Department of Energy, the Japanese 
Monbukagakusho, and the Max Planck Society. The SDSS Web site is 
{\tt http://www.sdss.org/}.

The SDSS is managed by the Astrophysical Research Consortium (ARC) for the 
Participating Institutions. The Participating Institutions are The 
University of Chicago, Fermilab, the Institute for Advanced Study, the 
Japan Participation Group, The Johns Hopkins University, the Korean 
Scientist Group, Los Alamos National Laboratory, the Max-Planck-Institute 
for Astronomy (MPIA), the Max-Planck-Institute for Astrophysics (MPA), New 
Mexico State University, University of Pittsburgh, University of 
Portsmouth, Princeton University, the United States Naval Observatory, and 
the University of Washington.

% \appendix

\end{document}